\documentstyle[12pt]{article}
\advance \topmargin by -\headheight
\advance \topmargin by -\headsep

\evensidemargin \oddsidemargin
\begin{document}

\topmargin -.6in

\newcommand{\sect}[1]{\setcounter{equation}{0}\section{#1}}
\renewcommand{\theequation}{\thesection.\arabic{equation}}

%
\def\rf#1{(\ref{eq:#1})}
\def\lab#1{\label{eq:#1}}
\def\nonu{\nonumber}
\def\br{\begin{eqnarray}}
\def\er{\end{eqnarray}}
\def\be{\begin{equation}}
\def\ee{\end{equation}}
\def\eq{\!\!\!\! &=& \!\!\!\! }
\def\foot#1{\footnotemark\footnotetext{#1}}
\def\lb{\lbrack}
\def\rb{\rbrack}
\def\llangle{\left\langle}
\def\rrangle{\right\rangle}
\def\blangle{\Bigl\langle}
\def\brangle{\Bigr\rangle}
\def\llbrack{\left\lbrack}
\def\rrbrack{\right\rbrack}
\def\lcurl{\left\{}
\def\rcurl{\right\}}
\def\({\left(}
\def\){\right)}
\newcommand{\nit}{\noindent}
\newcommand{\ct}[1]{\cite{#1}}
\newcommand{\bi}[1]{\bibitem{#1}}
\def\lskip{\vskip\baselineskip\vskip-\parskip\noindent}
\relax

\def\tr{\mathop{\rm tr}}
\def\Tr{\mathop{\rm Tr}}
\def\v{\vert}
\def\bv{\bigm\vert}
\def\Bgv{\;\Bigg\vert}
\def\bgv{\bigg\vert}
\newcommand\partder[2]{{{\partial {#1}}\over{\partial {#2}}}}
\newcommand\funcder[2]{{{\delta {#1}}\over{\delta {#2}}}}
\newcommand\Bil[2]{\Bigl\langle {#1} \Bigg\vert {#2} \Bigr\rangle}  
\newcommand\bil[2]{\left\langle {#1} \bigg\vert {#2} \right\rangle} 
\newcommand\me[2]{\left\langle {#1}\right|\left. {#2} \right\rangle} 
\newcommand\sbr[2]{\left\lbrack\,{#1}\, ,\,{#2}\,\right\rbrack}
\newcommand\pbr[2]{\{\,{#1}\, ,\,{#2}\,\}}
\newcommand\pbbr[2]{\lcurl\,{#1}\, ,\,{#2}\,\rcurl}
%
\def\a{\alpha}
\def\b{\beta}
\def\dc{{\cal D}}
\def\d{\delta}
\def\D{\Delta}
\def\eps{\epsilon}
\def\vareps{\varepsilon}
\def\g{\gamma}
\def\G{\Gamma}
\def\grad{\nabla}
\def\h{{1\over 2}}
\def\l{\lambda}
\def\L{\Lambda}
\def\m{\mu}
\def\n{\nu}
\def\o{\over}
\def\om{\omega}
\def\O{\Omega}
\def\p{\phi}
\def\P{\Phi}
\def\pa{\partial}
\def\pr{\prime}
\def\ra{\rightarrow}
\def\s{\sigma}
\def\S{\Sigma}
\def\t{\tau}
\def\th{\theta}
\def\Th{\Theta}
\def\ti{\tilde}
\def\wti{\widetilde}
\def\jc{J^C}
\def\bj{{\bar J}}
\def\sj{{\jmath}}
\def\bsj{{\bar \jmath}}
\def\bp{{\bar \p}}
\def\faa{Fa\'a di Bruno~}
\def\ca{{\cal A}}
\def\cb{{\cal B}}
\def\ce{{\cal E}}
\newcommand\sumi[1]{\sum_{#1}^{\infty}}   
\newcommand\fourmat[4]{\left(\begin{array}{cc}  
{#1} & {#2} \\ {#3} & {#4} \end{array} \right)}

%
\def\lie{{\cal G}}
\def\dlie{{\cal G}^{\ast}}
\def\elie{{\widetilde \lie}}
\def\edlie{{\elie}^{\ast}}
\def\hlie{{\cal H}}
\def\wlie{{\widetilde \lie}}
\def\f#1#2#3 {f^{#1#2}_{#3}}
\def\winf{{\sf w_\infty}}
\def\win1{{\sf w_{1+\infty}}}
\def\hwinf{{\sf {\hat w}_{\infty}}}
\def\Winf{{\sf W_\infty}}
\def\Win1{{\sf W_{1+\infty}}}
\def\hWinf{{\sf {\hat W}_{\infty}}}
\def\Rm#1#2{r(\vec{#1},\vec{#2})}          
\def\OR#1{{\cal O}(R_{#1})}           
\def\ORti{{\cal O}({\widetilde R})}           
\def\AdR#1{Ad_{R_{#1}}}              
\def\dAdR#1{Ad_{R_{#1}^{\ast}}}      
\def\adR#1{ad_{R_{#1}^{\ast}}}       
\def\KP{${\rm \, KP\,}$}                 
\def\KPl{${\rm \,KP}_{\ell}\,$}         
\def\KPo{${\rm \,KP}_{\ell = 0}\,$}         
\def\mKPa{${\rm \,KP}_{\ell = 1}\,$}    
\def\mKPb{${\rm \,KP}_{\ell = 2}\,$}    
%
\def\rlx{\relax\leavevmode}
\def\inbar{\vrule height1.5ex width.4pt depth0pt}
\def\IZ{\rlx\hbox{\sf Z\kern-.4em Z}}
\def\IR{\rlx\hbox{\rm I\kern-.18em R}}
\def\IC{\rlx\hbox{\,$\inbar\kern-.3em{\rm C}$}}
\def\one{\hbox{{1}\kern-.25em\hbox{l}}}
\def\0#1{\relax\ifmmode\mathaccent"7017{#1}%
        \else\accent23#1\relax\fi}
\def\omz{\0 \omega}
%
\def\ltimes{\mathrel{\vrule height1ex}\joinrel\mathrel\times}
\def\rtimes{\mathrel\times\joinrel\mathrel{\vrule height1ex}}
%
\def\mark{\noindent{\bf Remark.}\quad}
\def\prop{\noindent{\bf Proposition.}\quad}
\def\theor{\noindent{\bf Theorem.}\quad}
\def\name{\noindent{\bf Definition.}\quad}
\def\exam{\noindent{\bf Example.}\quad}
\def\proof{\noindent{\bf Proof.}\quad}
%
%
\def\PRL#1#2#3{{\sl Phys. Rev. Lett.} {\bf#1} (#2) #3}
\def\NPB#1#2#3{{\sl Nucl. Phys.} {\bf B#1} (#2) #3}
\def\NPBFS#1#2#3#4{{\sl Nucl. Phys.} {\bf B#2} [FS#1] (#3) #4}
\def\CMP#1#2#3{{\sl Commun. Math. Phys.} {\bf #1} (#2) #3}
\def\PRD#1#2#3{{\sl Phys. Rev.} {\bf D#1} (#2) #3}
\def\PLA#1#2#3{{\sl Phys. Lett.} {\bf #1A} (#2) #3}
\def\PLB#1#2#3{{\sl Phys. Lett.} {\bf #1B} (#2) #3}
\def\JMP#1#2#3{{\sl J. Math. Phys.} {\bf #1} (#2) #3}
\def\PTP#1#2#3{{\sl Prog. Theor. Phys.} {\bf #1} (#2) #3}
\def\SPTP#1#2#3{{\sl Suppl. Prog. Theor. Phys.} {\bf #1} (#2) #3}
\def\AoP#1#2#3{{\sl Ann. of Phys.} {\bf #1} (#2) #3}
\def\PNAS#1#2#3{{\sl Proc. Natl. Acad. Sci. USA} {\bf #1} (#2) #3}
\def\RMP#1#2#3{{\sl Rev. Mod. Phys.} {\bf #1} (#2) #3}
\def\PR#1#2#3{{\sl Phys. Reports} {\bf #1} (#2) #3}
\def\AoM#1#2#3{{\sl Ann. of Math.} {\bf #1} (#2) #3}
\def\UMN#1#2#3{{\sl Usp. Mat. Nauk} {\bf #1} (#2) #3}
\def\FAP#1#2#3{{\sl Funkt. Anal. Prilozheniya} {\bf #1} (#2) #3}
\def\FAaIA#1#2#3{{\sl Functional Analysis and Its Application} {\bf #1} (#2)
#3}
\def\BAMS#1#2#3{{\sl Bull. Am. Math. Soc.} {\bf #1} (#2) #3}
\def\TAMS#1#2#3{{\sl Trans. Am. Math. Soc.} {\bf #1} (#2) #3}
\def\InvM#1#2#3{{\sl Invent. Math.} {\bf #1} (#2) #3}
\def\LMP#1#2#3{{\sl Letters in Math. Phys.} {\bf #1} (#2) #3}
\def\IJMPA#1#2#3{{\sl Int. J. Mod. Phys.} {\bf A#1} (#2) #3}
\def\AdM#1#2#3{{\sl Advances in Math.} {\bf #1} (#2) #3}
\def\RMaP#1#2#3{{\sl Reports on Math. Phys.} {\bf #1} (#2) #3}
\def\IJM#1#2#3{{\sl Ill. J. Math.} {\bf #1} (#2) #3}
\def\APP#1#2#3{{\sl Acta Phys. Polon.} {\bf #1} (#2) #3}
\def\TMP#1#2#3{{\sl Theor. Mat. Phys.} {\bf #1} (#2) #3}
\def\JPA#1#2#3{{\sl J. Physics} {\bf A#1} (#2) #3}
\def\JSM#1#2#3{{\sl J. Soviet Math.} {\bf #1} (#2) #3}
\def\MPLA#1#2#3{{\sl Mod. Phys. Lett.} {\bf A#1} (#2) #3}
\def\JETP#1#2#3{{\sl Sov. Phys. JETP} {\bf #1} (#2) #3}
\def\JETPL#1#2#3{{\sl  Sov. Phys. JETP Lett.} {\bf #1} (#2) #3}
\def\PHSA#1#2#3{{\sl Physica} {\bf A#1} (#2) #3}
\def\PHSD#1#2#3{{\sl Physica} {\bf D#1} (#2) #3}
%
%

\begin{titlepage}
\vspace*{-1cm}
\noindent
April, 1993 \hfill{IFT-P/021/93}\\
\phantom{bla}
\hfill{UICHEP-TH/93-5} \\
\phantom{bla}
\hfill{hep-th/9304152}
\\
\vskip .3in

\begin{center}

{\large\bf On ${\bf W_{\infty}}$ Algebras, Gauge Equivalence of KP
Hierarchies,}
\end{center}
\begin{center}
{\large\bf Two-Boson Realizations and their KdV Reductions
\footnotemark
\footnotetext{Lectures presented at the VII J.A. Swieca Summer School,
Section: Particles and Fields, Campos do Jord\~ao - Brasil - January/93}}
\end{center}
\normalsize
\vskip .4in

\begin{center}
{ H. Aratyn\footnotemark
\footnotetext{Work supported in part by U.S. Department of Energy,
contract DE-FG02-84ER40173 and by NSF, grant no. INT-9015799}}

\par \vskip .1in \noindent
Department of Physics \\
University of Illinois at Chicago\\
845 W. Taylor St.\\
Chicago, Illinois 60607-7059\\
\par \vskip .3in

\end{center}

\begin{center}
{L.A. Ferreira\footnotemark
\footnotetext{Work supported in part by CNPq}}, J.F. Gomes$^{\,3}$,
and A.H. Zimerman$^{\,3}$

\par \vskip .1in \noindent
Instituto de F\'{\i}sica Te\'{o}rica-UNESP\\
Rua Pamplona 145\\
01405-900 S\~{a}o Paulo, Brazil
\par \vskip .3in

\end{center}

\begin{center}
{\large {\bf ABSTRACT}}\\
\end{center}
\par \vskip .3in \noindent

The gauge equivalence between basic KP hierarchies
is discussed. The first two Hamiltonian structures for KP hierarchies leading
to the linear and non-linear $\Winf$ algebras are derived. The realization
of the corresponding generators in terms of two boson currents is presented
and it is shown to be related to many integrable models which are
bi-Hamiltonian. We can also realize those generators by adding extra currents,
coupled in a particular way, allowing for instance a description of
multi-layered Benney equations or multi-component non-linear Schroedinger
equation. In this case we can have a second Hamiltonian bracket structure which
violates Jacobi identity.
We consider the reduction to one-boson systems leading to KdV and mKdV
hierarchies. A Miura transformation relating these two hierarchies is
obtained by restricting gauge transformation between corresponding two-boson
hierarchies.
Connection to Drinfeld-Sokolov approach is also discussed in the $SL(2,\IR)$
gauge theory.

\end{titlepage}
\sect{Introduction}

This is an expository account of results concerning
various integrable systems described as KP hierarchies and a method of
studying them by a symplectic gauge transformation.
The method allows to introduce an equivalence principle between
various KP hierarchies, which is useful in view of the growing number of
integrable models entering recently the area of high energy physics
(e.g. matrix models).
In the process one encounters different realizations of various
{\sf W}-infinity algebras in terms of two currents and new understanding
of KdV hierarchies and corresponding soliton equations via symplectic
reduction from two-boson KP hierarchies.

Section 2 introduces three basic KP hierarchies and the algebraic
structure behind their construction.
We point out that the Adler-Kostant-Symes (AKS) \ct{AKS}
theory with a Poisson bracket structure defined in terms of the R-matrix
\ct{STS83} Lie-Poisson bracket is the right setting to study these models
and to define the two-boson restriction of full KP hierarchy.
We also introduce the gauge transformation connecting basic flow equations.

In section 3 we present the hamiltonian structure of KP hierarchies
defining both first and second Poisson structures.
We also introduce here the fundamental two-boson KP hierarchy
used later in section 4 to construct various realization of the area
preserving  diffeomorphisms.
We go one step further in section 5 using two-boson KP hierarchy
to present two-current realization of $\Win1$ algebra and the corresponding
non-linear $\hWinf$ algebra.

Section 6 studies the concept of symplectic gauge equivalence between
various two-boson KP hierarchies playing the role of
generalized Miura transformations \ct{miura}.
We emphasize the symplectic character of equivalence of \mKPa and
${\rm KP}$ and show how this feature explains the 2-boson representation of
$\Win1$ and $\hWinf$ in terms of the \faa polynomials.

In Section 7 we apply Dirac reduction scheme to two-boson KP
hierarchies. We obtain in the process of reduction the standard
``one-boson" KdV and mKdV hierarchies. On reduced manifold the gauge
transformation connecting the two models takes the form of the Miura
transformation.

The results of section 6 and 7 can be rephrased using the language
of $ SL (2, \IR )$ gauge theory to reveal connection between
soliton equations and zero-curvature conditions. This is done in section 8.
This formalism enables us to formulate the Dirac reduction to KdV systems
in the language of Drinfeld-Sokolov reduction.

\sect{KP Hierarchies and their Algebraic Structure }
Let us introduce the following general KP pseudo differential operator
\ct{1,2,dickey}
\be
L \equiv  u_{-2} D + \sum_{i=-1}^{\infty} u_i D^{-i-1} = u_{-2} D + u_{-1} +
\sum_{i=0}^{\infty} u_i D^{-i-1}
\lab{1a}
\ee
where
\be
D \equiv {\pa \o \pa x}
\lab{1b}
\ee
and functions $u_i (x, t)$ dependent on infinitely many variables
$(x,t) = \(x, t_1, t_2, \ldots \)$.

In order to define $D^{-n}$, where $n$ is a positive integer, let us recall
a Leibniz rule
\be
D^{n} f = \sum_{\a = 0}^{\infty}{ n (n-1) \ldots (n- \a +1) \o \a !} \(
\pa^{\a}f\) D^{-n-\a}
\ee
and let $n \rightarrow -n$ obtaining:
\be
D^{-n} f = \sum_{\a = 0}^{\infty}(-1)^{\a}{(n+ \a -1)! \o \a ! (n-1)!} \(
\pa^{\a}f\) D^{-n-\a}
\lab{3}
\ee
Other useful relations are:
\br
\pa^{n} f &=& \sum_{\a =0}^{n} (-1)^{\a} {n \choose \a} D^{n-\a}f D^{\a}
\lab{4} \\
f D^{n} &=&\sum_{\a =0}^{n} (-1)^{\a}{n \choose \a}  D^{n-\a}\(\pa^{\a}f\)
\lab{5}
\er

There are three classes of integrable systems connected with the general
object $ L = u_{-2} D + u_{-1} + \sumi{i=0} u_i D^{-i-1}$ in \rf{1a}.
We label them by the parameter $\ell$ taking values $0,1,2$ and define
as follows \ct{GKR88,BAK85,oevelr,kiso,ANPV}:

\nit $\ell=0\,\;{\bf :} \;$ Lax operator $L$ with $u_{-2} =1$ and $u_{-1} =0$
(standard KP case)

\nit $\ell=1\,\;{\bf :} \;$ Lax operator $L$ with $u_{-2} =1$ and
$u_{-1} \ne 0$ (first non-standard  KP case)

\nit $\ell=2\,\;{\bf :} \;$ Lax operator $L$ with the most general form as in
\rf{1a} (second non-standard KP case)

It appears that existence of these three classes has origin in fundamental
algebraic properties of a group of pseudo-differential operators on a circle.
We will extract here few basic details of the algebraic formalism
\ct{GKR88,BAK85,ANPV}.
An object of interest is the Lie algebra $\lie $
of pseudo-differential operators on a circle. An element
of $\lie$ is given by an arbitrary
pseudo-differential operator $X = \sum_{k \geq -\infty}
D^k \,X_k (x) \,$.

Origin of three models presented above can be traced to the fundamental
fact that there exist precisely three decompositions of $\lie$ into a linear
sum of two subalgebras \ct{GKR88,BAK85,ANPV} (i.e. $\lie = \lie_{+}^{\ell}
\oplus \lie_{-}^{\ell}$, parametrized
by the same index $\ell$ taking values $\ell = 0,1,2$):
\be
\lie^{\ell}_{+} = \{\, X_{\geq \ell}
= \sumi{i=\ell} D^i X_i (x) \,\}
\quad;\quad
\lie^{\ell}_{-} = \{\, X_{< \ell}
= \sumi{i=-\ell+1 } D^{-i} X_{-i}(x)\,\}
\lab{subalg}
\ee
The dual spaces to subalgebras $\lie^{\ell}_{\pm}$ are given
by:
\be
{\lie^{\ell}_{+}}^{\ast} = \{ L_{< -\ell}
= \sumi{i=\ell+1} u_{-i}(x) D^{-i}\, \} \;\; ;\;\;
{\lie^{\ell}_{-}}^{\ast} = \{  L_{\geq -\ell} =
\sumi{i=-\ell} u_{i}(x) D^{i}\}
\lab{dsubalg}
\ee
Application of the general Adler-Kostant-Symes (AKS) \ct{AKS} formalism
results in all three decompositions giving rise to integrable models with
flow equations allowing generalized Lax representations:
\be
\partder{L}{t_n} = \sbr{P_{\ge \ell} (L^n)}{L}
\lab{aksflow}
\ee
where the projection $P_{\ge \ell}$ projects on terms $a_i D^i$ with
$i \ge \ell$.

The basic structure of AKS construction is
the R-matrix, which for all the above cases is defined as
$R_{\ell} \equiv P_{+}^{\ell}  - P_{-}^{\ell} $, where
$P_{\pm}^{\ell}$ are projections on $\lie_{\pm}^{\ell}$.
It follows from the general formalism that
\be
\lb X , Y \rb_{R_{\ell}}
\equiv \h \, \lb R_{\ell} X , Y \rb + \h \, \lb X, R_{\ell} Y \rb
=\lb X_{\geq \ell}, Y_{\geq \ell}\rb - \lb X_{< \ell} , Y_{< \ell} \rb
\lab{r-brack}
\ee
defines an additional (with respect to usual commutator)
Lie structure on $\lie\,$ (see \ct{ANPV} and references therein).
{}From the general relation for the $R$-coadjoint action of
$\lie\,$ on its dual space we find
that the infinitesimal shift along an $R$-coadjoint orbit
$\OR{\ell}\,$ has the form: $\d_{R_\ell} L \equiv
ad^{\ast}_{R_\ell} (X) L =
\lb X_{\geq \ell} , L_{< -\ell} \rb_{< -\ell}
- \lb X_{< \ell} , L_{\geq -\ell} \rb_{\geq -\ell}$.
We now make contact with three special cases of the general Lax from \rf{1a}
showing how they appear naturally in the AKS formalism sketched above.

\subsection{Standard \KPo Hierarchy. Sato's Theorem.}
The \KPo model is defined in terms of the element of coadjoint orbit.
The construction goes as follows. Note that $D \in {\lie^{\ell=0}_{+}}^{\ast}$
is invariant under the coadjoint action $\d_{R_{\ell=0}} D=0$.
By adding to it the general elements of ${\lie^{\ell=0}_{-}}^{\ast}$ from
\rf{dsubalg} we arrive at the $R$-coadjoint orbit of the form
$\, \OR{0} = \{ L \}$ with
\be
L \equiv D +  \sumi{i=0} u_i (x, t) D^{-1-i}
\lab{laxop}
\ee
According to \rf{aksflow} evolution of this standard
KP hierarchy is governed by flows equations
\be
\partder{L}{t_n} = \lb B_n^{-} \, , \, L \rb
= \lb B_n^{+} \, , \, L \rb \qquad\;\; n=1, 2, \ldots
\lab{kpflow}
\ee
where we have introduced the potentials:
\be
B_n^{-} \equiv - \(L^n \)_{-} \qquad ; \qquad
B_n^{+} \equiv \(L^n \)_{+}
\lab{bnplus}
\ee
where the subscript $(+)$ means taking the purely differential part of the
corresponding operator.

One can establish by induction that
\be
\partder{L^m}{t_n} = \lb B_n^{-} \, , \, L^m \rb
\lab{flowlm}
\ee
holds too.
As a consequence we find $\pa B_m^{-}/ \pa t_n = \(\lb B_n^{+} \, , \,
B_m^{-} \rb \)_{-}$ from which one derives the Zakharov-Shabat equations:
\be
\partder{B_m^{-}}{t_n} - \partder{B_n^{-}}{t_m}
 = \lb B_n^{-} \, , \, B_m^{-} \rb
\lab{zseqs}
\ee
The following fundamental result is due to Sato:

\theor There exists a pseudo-differential operator  $W$
\be
W \equiv 1 + \sumi{i=1}  w_i (x,t) D^{-i}
\lab{satow}
\ee
for which the following Sato equations are valid:
\br
L \eq W D W^{-1}   \lab{satoeqs} \\
\partder{W}{t_n} \eq  B_n^{-} W = B_n^{+} W - W D^n \nonu
\er

\subsection{Nonstandard Hierarchies. Gauge Map to the KP Hierarchy.}

We now turn our attention to \mKPa hierarchy.
We first consider elements of
${{\cal G}_{-}^{\ell=1}}^{\ast} $ of the type $L^{(1)} = D + u_{-1} +
\, u_0 D^{-1} $, which
preserve their form under $\d_{R_1} L^{(1)} = ad^{\ast}_{R_1} (X) L^{(1)}$,
spanning therefore a finite $R_1$-orbit.
A complete Lax operator is obtained by adding $L^{(1)}$ to the general
element $L_{-}$ of ${{\cal G}_{+}^{\ell=1}}^{\ast} $ obtaining
\be
L= L^{(1)} + L_{-} = D + u_{-1} +
\, u_0 D^{-1} + \sum_{i \geq 1} u_{i}\,D^{-i-1}.
\lab{mkpalax}
\ee

According to \rf{aksflow} this Lax operator satisfies:
\be
\partder{L}{t_n} = \lb P_{\ge 1} (L^n) \, , \, L \rb
\lab{mkpflow}
\ee
As pointed out already by Sato (as referenced in \ct{kiso}) there is a gauge
transformation :
\be
K \equiv G^{-1} L G  = D +  \sumi{i=0} v_i D^{-1-i} \qquad; \qquad
G \equiv \exp \( - \int^x u_{-1}\, dx \)
\lab{gaug1}
\ee
which removes the constant $u_{-1}$ term and gives rise to the transformed
Lax operator of the standard \KPo form.
In fact, this gauge transformed Lax operator satisfies the standard
KP flow equation as well.
This is a basic result, described by Kiso in \ct{kiso}. The argument goes as
follows.

\theor If $L$ satisfies the \mKPa flow equation \rf{mkpflow} then
the gauge transformed Lax $K$ satisfies the standard KP evolution equation
\rf{kpflow}: $ \pa K / \pa t_n = \sbr{(K^n)_{+}}{K}$.

{}From definition and \rf{mkpflow} we find by direct calculation and use of
\rf{mkpflow} that:
\be
\partder{K}{t_n} = \partder{G^{-1} L G}{t_n} =
\left\lb G^{-1} P_{\ge 1} (L^n) G - G^{-1} \partder{G}{t_n} \, , \, K
\right\rb
\lab{kt1}
\ee
Using the definition $ P_{\ge 1} (L^n) = L^n - \sum_{j \le 0} (L^n )_j$
we find $ \sbr{P_{\ge 1} (L^n) }{L}=
- \sbr{\sum_{j \le 0} (L^n )_j}{L}$ and consequently
\be
\partder{u_{-1}}{t_n} = \( \partder{L}{t_n}\)_0 =
- \( \sbr{\sum_{j \le 0} (L^n )_j}{L}\)_0 = \pa L^n_0
\lab{kt2}
\ee
This last relation leads to
\be
\partder{G}{t_n} = - \int^x \partder{u_{-1}}{t_n} dx \; G=
- \int^x \pa L^n_0 dx \; G = - L^n_0 \; G
\lab{kt3}
\ee
Hence, we obtain
\br
G^{-1} P_{\ge 1} (L^n) G - G^{-1} \partder{G}{t_n}
\eq G^{-1} \(  L^n - \sum_{j \le 0} (L^n )_j\)    G - G^{-1} \partder{G}{t_n}
\nonu \\
\eq K^n - G^{-1} \sum_{j \le -1 } (L^n )_j  G  = \( K^n \)_+
\lab{kt4}
\er
Inserting this in \rf{kt1} we arrive at the standard KP evolution equation
\rf{kpflow}.

Finally, let us make few comments on the general case of $\ell=2$
of \mKPb hierarchy.
Here elements of ${{\cal G}_{-}^2}^{\ast} $
of the form $L^{(2)} = u_{-2} D + u_{-1} + u_0 D^{-1} +
u_1 D^{-2} $ span an invariant subspace under
$\d_{R_2} L^{(2)} = ad^{\ast}_{R_2} (X) L^{(2)} $.
Defining the complete Lax operator as $L = L^{(2)} + L_{-} =
u_{-2} D + u_{-1} + u_0 D^{-1} + u_1 D^{-2} +
\sum_{i \ge 2} u_i D^{-i-1} $ we find \rf{1a} in its most general form.

\sect{Hamiltonian Structure of KP Hierarchies}

\subsection{The First Hamiltonian Structure of KP Hierarchies}

Let us go back to the definitions \rf{subalg} and \rf{dsubalg} of relevant
subspaces of $\lie$ and $\dlie$.
Using the Adler trace one defines an invariant, non-degenerate
bilinear form:
\be
\blangle L \v X \brangle \equiv {\Tr} \( L X \) =
\int dx \; \mbox{\rm Res} \( L X\)
\lab{6}
\ee
where the residue (Res) means the coefficient of $D^{-1}$. The above $\Tr$
has the cyclic property.

We focus on a case of \mKPa with the Lax operator \rf{mkpalax}.
Let
\be
Q = \sum_{n=-1}^{\infty} D^{n} q_{n} \qquad \, ; \qquad \,
V = \sum_{n=-1}^{\infty} D^{n} v_{n}
\lab{7}
\ee
be objects in $\lie = \lie_{+}^{\ell=1} \oplus
\lie_{-}^{\ell=1}$ dual to $L$ from \rf{mkpalax}
(the fact that $Q,V$ are truncated algebra
elements will not alter the generality of our discussion).
Inserting $Q$ into the definition \rf{6} yields
\be
\langle L \mid Q \rangle = \sumi{n=-1} \int dx\; u_n(x) q_n(x)
\lab{8}
\ee
as can easily be seen from
\be
\langle L \mid Q \rangle = \sum_{n=-1}^{\infty}\sum_{i=-1}^{\infty}
\int dx\; \mbox{\rm Res} \, \, \( u_i(x) D^{-i-1+n}q_n(x)\)
\lab{9}
\ee
which, because of \rf{3}, gets the contribution from $i=n$ terms
only.

Recall now the definition of R-bracket given in \rf{r-brack} for the $\ell=1$
case.
It turns out that the first Hamiltonian structure coincides with
a first Gelfand-Dickey \ct{dickey,GD} bracket:
\be
\{ \langle L \mid Q \rangle \, , \, \langle L \mid V \rangle \}_1^{\ell=1}
 \equiv \langle L \mid \lb Q \, , \, V \rb_{\ell=1} \,  \rangle
\lab{10}
\ee
with
\be
\lb Q \, , \, V \rb_{\ell=1} \equiv
\sum_{n,m=1}^{\infty} \lb D^n q_n \, , \, D^m v_m \rb -
\lb q_0 + D^{-1} q_{-1} \, , \, v_0+ D^{-1} v_{-1} \rb
\lab{11}
\ee
Using \rf{5} it follows that the first term in \rf{11} is given by
\be
\sum_{n,m=1}^{\infty}\sum_{\a=0}^{m} (-1)^{\a} {m \choose \a} D^{n+m-\a}
q_n^{(\a )} v_m -
\sum_{n,m=1}^{\infty}\sum_{\a=0}^{m} (-1)^{\a} {n \choose \a} D^{n+m-\a}
q_n v_m^{(\a )}
\lab{12}
\ee
with
\be
q_n^{(\a )} \equiv  {\pa^{\a} q_n \o \pa x^{\a}} \, \, \, ; \, \, \,
v_n^{(\a )} \equiv  {\pa^{\a} v_n \o \pa x^{\a}}
\lab{13}
\ee
while the relevant terms from the second term in \rf{11} are
$ - v_{-1} q_0^{\pr} D^{-2} +v_{0}^{\pr} q_{-1} D^{-2} $.
The calculation of $\langle L \mid \lb Q , V \rb_{\ell=1} \rangle$
gives after integration by parts:
\br
\langle L \mid \lb Q , V \rb_{\ell=1} \rangle &=& \sumi{n,m=0} \int dx dy
\sum_{\a =0}^{m} {m \choose \a} \pa^{\a}_x \( u_{n+m-\a}(x) \d (x-y) \)
 q_n(x) v_m(y) -
\lab{14}\\
& & \sum_{n,m=0}^{\infty} \int dx dy \sum_{\a =0}^{n} {n \choose \a}(-1)^{\a}
  u_{n+m-\a}(x) \( \pa^{\a}_x \d (x-y) \)
 q_n(x) v_m(y)
\nonumber \\
&+& \int dx dy (\d^{\pr} (x -y) ) \( v_{-1} (y) q_0 (x)
- v_{0} (y) q_{-1}(x) \) \nonu
\er
Using \rf{8} in the l.r.s. of \rf{10} and comparing the coefficients of
$q_n(x) v_m(y)$ we obtain from \rf{14}
\br
\{ {\ti u}_n(x)\, , \, {\ti u}_m(y) \}_1^{\ell =1} \eq
\{ u_{n+1} (x)\, , \, u_{m+1} (y) \}_1^{\ell =1} =
\Omega^{(1)}_{nm}({\ti u}(x)) \d (x-y)
\quad n,m =0, 1,\ldots \phantom{aaa}
\lab{15a}\\
\{ u_{0}(x)\, , \, u_{m }(y) \}_1^{\ell =1} \eq
\{ u_{-1}(x)\, , \, u_m(y) \}_1^{\ell =1} = 0
\lab{15b}\\
\{ u_{0}(x)\, , \, u_{-1}(y) \}_1^{\ell=1} \eq \d^{\pr} (x -y) \; ;\;
\{ u_{0}(x)\, , \, u_{0}(y) \}_1^{\ell=1}
= \{ u_{-1}(x)\, , \, u_{-1}(y) \}_1^{\ell=1} =0
\lab{15c}
\er
where we have introduced for convenience ${\ti u}_n(x) = u_{n+1} (x)$
and where
\be
\Omega^{(\ell)}_{nm}(u(x)) \equiv - \sum_{k=0}^{n+\ell}(-1)^k
{n+ \ell\choose k} u_{n+m+\ell-k}(x) D^k_x +
\sum_{k=0}^{m+\ell} {m+\ell\choose k} D^k_x
u_{n+m+\ell-k}(x)
\lab{omega}
\ee
We see therefore that the pair $\( u_0, u_{-1} \)$ decouples from the
rest of algebra. This result will soon become very crucial for our discussion
of two-boson systems.

Let us now turn our attention to the case of standard KP hierarchy with
$\ell=0$. Here construction of the first Gelfand-Dickey bracket
is based on the commutator:
\be
\lb Q \, ,\, V \rb_{\ell=0}  \equiv
\sum_{n,m=0}^{\infty} \lb D^n q_n \, , \, D^m v_m \rb -
\lb D^{-1} q_{-1} \, , \, D^{-1} v_{-1} \rb
\lab{11a}
\ee
Using \rf{5} it follows that the first term in \rf{11a} is given by
\be
\sum_{n,m=0}^{\infty}\sum_{\a=0}^{m} (-1)^{\a} {m \choose \a} D^{n+m-\a}
q_n^{(\a )} v_m -
\sum_{n,m=0}^{\infty}\sum_{\a=0}^{m} (-1)^{\a} {n \choose \a} D^{n+m-\a}
q_n v_m^{(\a )}
\lab{12a}
\ee
We can now calculate $\langle L \mid \lb Q , V \rb_{\ell=0} \rangle$.
Although in this case one should consequently take $L$ Lax operator as given
by \rf{laxop} we will keep using $L$ from \rf{mkpalax} anticipating that
anyway the extra term $u_{-1}$ will automatically decouple from the
remaining variables.
This time only the first
term of \rf{11a} contributes, giving after integration by parts:
\br
\langle L \mid \lb Q , V \rb_{\ell=0} \rangle &=& \sumi{n,m=0} \int dx dy
\sum_{\a =0}^{m} {m \choose \a} \pa^{\a}_x \( u_{n+m-\a}(x) \d (x-y) \)
 q_n(x) v_m(y) -
\nonumber \\
& & \sum_{n,m=0}^{\infty} \int dx dy \sum_{\a =0}^{n} {n \choose \a}(-1)^{\a}
  u_{n+m-\a}(x) \( \pa^{\a}_x \d (x-y) \)
 q_n(x) v_m(y)\nonumber\\
& &\lab{14a}
\er
Using \rf{8} in the l.r.s. of \rf{10} and comparing the coefficients of
$q_n(x) v_m(y)$ we obtain from \rf{14a}
\br
\{ u_n(x)\, , \, u_m(y) \}_1^{\ell=0} &=& \Omega^{(0)}_{nm}(u(x)) \, \d (x-y)
\lab{15aa}\\
\{ u_{-1}(x)\, , \, u_{-1}(y) \}_1^{\ell=0} &=&
\{ u_{-1}(x)\, , \, u_m(y) \}_1^{\ell=0} = 0
\lab{15ba}
\er
We see therefore that $u_{-1}$, in fact decouples from the rest of the
algebra \rf{15aa}.
The algebra \rf{15aa} was first derived by Watanabe \ct{watanabe}
(see also \ct{BAK85,wu91}).
Let us note that for $n=m=1$, \rf{15aa} gives the Virasoro algebra without
central term. By measuring the spin of the fields $u_n(x)$ through the
Virasoro field $u_1(x)$ (which has spin $1$) we see that $u_0(x)$ has
spin $2$, $u_2(x)$ has spin $3$ and so on.
Therefore \rf{15aa} describes the $\Win1$ algebra
without central term.

\subsection{The Second Hamiltonian Structure}
The second Gelfand-Dickey \ct{dickey} bracket is given by
\be
\{\langle L\mid Q \rangle \, , \, \langle L\mid V \rangle\}_{2} =
\Tr\(L(QL)_{+}V -  (LQ)_{+}LV\) = \Tr\((LQ)_{-}LV - L(QL)_{-}V\)
\lab{16}
\ee
where the subscripts $\pm$ denote the parts of the pseudo-differential operator
containing non-negative and negative powers of $D$ and where we continue to
use the Lax operator $L$ from \rf{mkpalax}.

By performing explicit calculation of \rf{16} we obtain \ct{1,2}
\br
\{ u_n (x) \, , \, u_m (y) \}^{GD}_2
&=&\! \Omega^{(1)}_{nm}(u(x))\d (x-y) \lab{17a}\\
&+&\! \sum_{i=0}^{m-1} \Bigg\lb \sum_{k=1}^{m-i-1} {m-i-1 \choose k} u_i (x)
D^k_x u_{m+n-i-k-1} (x) \nonumber \\
&-&\! \sum_{k=1}^{n} (-1)^k {n \choose k} u_{n+i-k} (x) D^k_x
u_{m-i-1} (x) \Bigg\rb \, \d (x-y)  \nonumber\\
-\sum_{i=0}^{m-1} \sum_{k=0}^{n}\sum_{l =1}^{m-i-1} \!\!\! &(-1)^k &\!\!\!
{n \choose k} {m-i-1 \choose l} u_{n+i-k}(x) D^{k+l}_x
u_{m-i-l-1} (x) \d (x-y)
\nonumber
\er
for $m,n \geq 0$ together  with
\br
\{ u_{n}(x),u_{-1}(y)\}_{2} &=& -\sum_{k=1}^{n}(-1)^{k}{n \choose
 k}u_{n-k}(x)D_{x}^{k}\d (x-y)
\lab{17b}\\
\{ u_{-1}(x),u_{m}(y)\}_{2} &=&
\sum_{k=1}^{m}{m \choose k}D_{x}^{k}u_{m-k}(x)\d (x-y)
\lab{17c}\\
\{ u_{-1}(x),u_{-1}(y)\}_{2} &=& - \d^{\pr} (x-y)
\lab{17d}
\er
We then see that $u_{-1}(x)$ couples to itself and to other fields $u_{n}(x)$
for $n\neq 0$.
Recall however that by the appropriate gauge transformation \rf{gaug1}
we were able to remove the constant $u_{-1}$ term
casting the transformed Lax operator into the standard \KPo form.
One suspects therefore that one can always impose the condition
$u_{-1} = 0$.
In view of \rf{17d}  $u_{-1} = 0$ represents a second
class  constraint and thus we consider the Dirac bracket \ct{1}:
\br \lefteqn{
\{ u_{n} (x) , u_{m} (y) \}_{2}^{D} =\{ u_{n}(x), u_{m}(y)\}_{2}}\lab{19}\\
&-&\int \int
dz_{1}dz_{2} \{ u_{n}(x), u_{-1}(z_{1})\}_{2}
\{ u_{-1}(z_1),u_{-1}(z_2)\}_{2}^{-1} \{ u_{-1}(z_2),u_{m}(y)\}_{2}
\nonumber
\er
which leads, with the use of eqs. \rf{17a},\rf{17b},\rf{17c},\rf{17d} to
\be
 \{ u_{n}(x), u_{m}(y)\}_{2}^{D} =\{ u_{n}(x), u_{m}(y)\}_{2} -
\sum_{i=0}^{n-1}\sum_{j=0}^{m-1}(-1)^{n-i} {n \choose i}{m \choose j}
u_{i}(x)D_{x}^{m+n-i-j-1}u_{j}(x)\d (x-y)
\lab{20}
\ee
We recognize in the second term on the r.h.s of \rf{20} a
Drinfeld-Sokolov bracket.
Notice that it satisfies  the Jacobi
identity due to the fact that \rf{16} or \rf{17a} also do it.
The algebra in \rf{20} describes a nonlinear $\hWinf$ algebra since
its lowest spin generator $u_0$ satisfies the Virasoro algebra.

Let us make a comment about the quasiclassical counterparts of the algebraic
structures we studied above.
Making in \rf{15aa} the substitution $\pa_x \rightarrow h\pa_x$ and setting
$h \rightarrow 0$, we obtain the ``classical limit''  of $\Win1$ algebra:
\be
\{\omega_n(x), \omega_m(y)\}_{1} = \( n \, \omega_{m+n-1}(x)D_x +
 m\, D_{x}\omega_{m+n-1}(x) \)\, \d (x-y)
\lab{27}
\ee
which is the area preserving diffeomorphism algebra called $\win1$
algebra.
Using the same trick in equation \rf{20} we obtain the non-linear $\hwinf$
algebra, namely:
\br
\{\omega_n(x), \omega_m(y)\}_{1} \eq \( (n+1)\omega_{m+n}(x)D_x + (m+1)
D_{x}\, \omega_{m+n}(x) \) \, \d (x-y) \nonu \\
\!\!&+&\!\! 2nm \, \om _{n-1}(x)D_x\om _{m-1}(x) \,\d (x-y)
\lab{27l}
\er
\subsection{Hamiltonian Structure and KP equation}

The first  bracket \rf{15aa} and the second bracket \rf{20} lead to an
Hamiltonian description of the KP flow equation \rf{kpflow}.
The Hamiltonians are defined in terms of $L$ from \rf{laxop} as
\be
H_m = {1 \o m}\, \Tr\, L^m = {1 \o m}\, \int {\rm Res}\, L^m
\lab{21}
\ee
and satisfy
\be
{{\pa L \o {\pa t_m}}}= [(L^m)_{+}, L] = \{L,H_{m+1}\}_{1} = \{L,H_m\}_2^{D}
\lab{22}
\ee
Equivalently, we have
\be
{{\pa u_n \o {\pa t_m}}}=  \{ u_n,H_{m+1}\}_{1} = \{u_n,H_m\}_2^{D}
\lab{23}
\ee
{}From \rf{21} we find:
\br
H_1 \eq \int u_0(x)\,dx \quad ;\quad H_2 =\int u_1(x)\,dx     \nonumber\\
H_3 \eq \int (u_2(x)+ u_0^2)\, dx \quad ;\quad H_4 =\int
(u_3(x)+3 u_1(x) u_0(x))\, dx \nonumber\\
H_5 \eq \int \( u_4 +4 u_0 u_2 + 2 u_1^2 +2 u_0^3 + u_0 u_0^{\pr \pr} -
2 u_1 u_0^{\pr}\) \, dx
\lab{24}
\er
Expressions \rf{23} and \rf{24} give after the use of \rf{15a} and
\rf{20} \ct{2}:
\br
 {\pa u_n \o{\pa x}}&=&{{\pa u_n }\o {\pa t_1}}=  \{ u_n,H_{2}\}_{1} =
\{ u_n,H_1\}_2^{D} \nonumber\\
{{\pa u_n} \o {\pa y}}&=&{{\pa u_n }\o {\pa t_2}}=  \{ u_n,H_{3}\}_{1} = \{
 u_n,H_2\}_2^{D}= {\pa ^2u_n \o {
{\pa x}^2}} + 2 {\pa u_{n+1} \o {\pa x}}  -2 \sum_{k=1}^{n}(-1)^k {n \choose
k}u_{n-k}{\pa^k u_0 \o {\pa x}^k} \nonumber\\
{{\pa u_0 }\o {\pa t}}&=&{{\pa u_0} \o {\pa t_3}}= \{ u_0,H_{4}\}_{1} =
\{ u_0,H_3\}_2^{D}= {\pa^3u_0 \o {{\pa x}^3}} + 6u_0{\pa u_0 \o {\pa x}} +
3({\pa^2u_1 \o {\pa x}^2} + {\pa u_2 \o {\pa x}})
\lab{25}
\er
where we have denoted $t_1 =x$, $t_2 = y$ and $t_3 = t$.  Using the first two
equations we obtain from the third  equation in \rf{25} :
\be
{\pa \o {\pa x}}({\pa u_0 \o {\pa t}}-{1 \o 4}{\pa^3u_0 \o {\pa x}^3}  - 3 u_0
{\pa u_0 \o \pa x}) = {3 \o 4}{\pa^2u_0 \o {\pa y}^2}
\lab{26}
\ee
which is the KP (Kadomtsev-Petviashvili) equation.  Let us mention that if we
constraint our Lax operator \rf{1a} by setting $u_1 = u_2=...=0$, then
the third eqn \rf{25} gives the KdV equation for $u_0$.  This  will be more
explicitly seen in terms of the two current realization of the KP-hierarchy.

\subsection{Two-Boson KP Hierarchy}
Here we go back to \mKPa and make the following crucial observation.
Consider truncated elements of
${{\cal G}_{-}^1}^{\ast} $ of the type $L_{J}^{(1)} = D + u_0 +  \,u_1\,
D^{-1} = D - J + \bj D^{-1} $, where we have introduced two Bose currents
$(J,\bj)$ to create fit with notation used in \ct{2boson}.
Recall that under the coadjoint action $\d_{R_1} L_{J}^{(1)} =
ad^{\ast}_{R_1} (X) L_{J}^{(1)}$ this finite Lax operator maintains its form,
i.e. the two-boson Lax operators span an $R_1$-orbit of finite functional
dimension $\, 2$.
This observation, already present in \ct{GKR88} clarifies status of
two-boson $(J,\bj)$ system as a consistent restriction of the full \mKPa
hierarchy understood as an orbit model.
Note that there are only two possibilities for the invariant $R_1$-orbit;
the two-boson system and the full \mKPa system (in a quasiclassical limit
situation is much richer with any number of fields defining invariant
subspace).
A Poisson bracket obtained as Lie-Poisson $R$-bracket in \rf{15c} yields
the first bracket structure of the two-boson $(J,\bj)$ system:
\br
\{\bj (x),J(y)\}_{1} &=& -\d^{\prime}(x-y) \nonumber \\
\{ J(x),J(y)\}_{1} &=& \{\bj (x),\bj (y)\}_{1} = 0
\lab{29}
\er

The higher bracket structures have been investigated in \ct{BAK85,2boson}.
One finds the following second bracket structure:
\br
\{\bar J(x),J(y)\}_2 &=& J(x) \d^{\prime}(x-y) - h \d^{\pr \pr} (x-y) \nonu\\
\{\bar J(x),\bar J(y)\}_2 &=& 2 \bar J(x)\d^{\prime}(x-y) +
\bar J^{\prime}(x)\d (x-y)\nonu \\
\{J(x),J(y)\}_2 &=& c\d^{\prime}(x-y)
\lab{33}
\er
Consistency check based on Lenard relation forces
the deformation parameters $c,h$ to take values $c=2\,,\,h=1$.

The three lowest Hamiltonian functions are:
\be
H_{J\,1} = \int \bj \quad;\quad H_{J \,2} = \int - \bj J
\quad;\quad
H_{J\,3} = \int \( \bj J^2 + \bj J^{\pr} + \bj^2 \) \lab{hJ}
\ee
For the general Hamiltonian matrix structure $P_i$ we have
\be
\partder{}{t_r} { J \choose \bj} = P_{J\,i} \,
{ {\d H_{J\, r+2-i}}/ {\d J} \choose
{\d H_{J\,r+2-i}}/ {\d \bj}} = P_{J\,1} {\d H_{r+1}/\d J\choose
\d H_{r+1}/ \d \bj } = P_{J\,2} { \d H_{r}/ \d J \choose
\d H_{r}/ \d \bj}
\lab{iflow}
\ee
Among the multi-Hamiltonian structures only $P_{J\,1}$ and $P_{J\,2}$ are
independent.
All other matrices $P_{J\,i}\;,\; i=3,4,\ldots$ are related
to $P_{J\,2}$ through $P_{J\,i} =\(P_{J\,2} (P_{J\,1})^{-1} \)^{i-2} P_{J\,2}$
involving the so-called recurrence matrix $P_{J\,2} (P_{J\,1})^{-1} $
\ct{fordy,2boson}.
The explicit form of first and second local Hamiltonian structures
corresponding to \rf{29} and \rf{33} with $c=2$ and $h=1$ is:
\be
P_{J\,1} = \left(\begin{array}{cc}
0 & - D \\
-D & \; 0 \end{array}
\right) \;\; , \;\;
P_{J\, 2} =\left(\begin{array}{cc}
2 D & \; D^2 + D J \\
- D^2 + J D &\; D \bj+ \bj D\end{array}
\right)
\lab{Jp1Jp2}
\ee
Taking $r=2$ in \rf{iflow} we especially get the Boussinesq equation:
\br
J_{t_2} \eq {\pbr{J}{H_{J\,2}}}_2 = {\pbr{J}{H_{J\,3}}}_1
-h J^{\pr \pr} - \( J^2 \)^{\pr} - 2
\bj^{\pr}  \nonu\\
\bj_{t_2} \eq {\pbr{\bj}{H_{J\,2}}}_2 = {\pbr{\bj}{H_{J\,3}}}_1
h \bj^{\pr \pr} - 2 \(\bj J\)^{\pr}
\lab{boussin}
\er
where we re-introduced $h$ as a deformation parameter.
In the dispersiveless limit $h \to 0$ taken in \rf{boussin}
we obtain the classical dispersiveless long wave equations (Benney
equations) \ct{LM79,BAK85}.

\sect{Realization of $\omega_{1+\infty}$  and ${\hat \omega}_{\infty}$ in
 Terms of Currents - Applications to Benney Equations and WZNW}
\subsection{The Benney Equations.}
Consider the following realization of the area-preserving generators:
\be
\omega_n = (-1)^n \bar J(x)J(x)^n
\lab{28}
\ee
$n=0,1,2,...$ where the currents $J$ and $\bj$ satisfy the first bracket
structure \rf{29}.
It is not difficult to verify that \rf{28} satisfy \rf{27} after use is made of
\rf{29}.

Taking $H_3$
\be
H_3 = \int (\omega_2(y) + \omega_0^2(y))dy = \int (\bar J(y)J^2(y) +
\bar J^2(y))dy
\lab{31}
\ee
as the evolution generator in time for our system with respect to
the first bracket, yields:
\br
{dJ(x) \o {dt}} &=& {dJ(x) \o {dt_2}} = \{J(x),H_3\}_1=
- (J^2(x))^{\pr} - 2 \bar J^{\pr}(x) \nonu\\
{d\bar J(x) \o {dt}} &=& {d\bar J(x) \o {dt_2}} =\{\bar J(x),H_3\}_1
= -2 (\bar J(x) J(x))^{\pr}
\lab{30}
\er
where $J^{\prime}(x) = {dJ(x) \o {dx}}$, etc.

The non-linear deformation of the area preserving diffeomorphism algebra
\rf{27l} can be generated by \rf{28} where now $J$ and $\bar J$ satisfy the
second-bracket structure \rf{33} with the deformation parameter $h$ set to
zero.
Recalling \rf{23} it follows using \rf{33} that
\br
{dJ(x) \o {dt}}& =&\{J(x),H_2\}_2 = - (J^2(x))^{\prime} - 2 \bar J^{\prime}(x)
\nonu\\
{d\bar J(x) \o {dt}} &=& \{\bar J(x),H_2\} =-2 (\bar J(x) J(x))^{\prime}
\lab{34}
\er
where $H_2 = -\int \bar J(y)J(y)dy$.  Equations \rf{30} or \rf{34} are known
to be the Benney equations of hydrodynamics \ct{LM79,ZA,BAK90}.

In this way the Benney equations are related to the classical KP-hierarchy
where the first Poisson bracket structure is defined by $\win1$
while the second bracket is given by the non-linear extension of $\winf$.

In an attempt to generalize \rf{28} let us consider $N$ copies of
currents $J$ and $\bar J$ and define
\be
\omega_n (x) = (-1)^n \sum_{k=1}^{N} \bar J_k(x) (J_k(x))^n
\lab{35}
\ee
The Lax equations for $\om_0$ and $\om_1$ with respect to $t_2$ and $t_3$
take the form:
\br
{d\om _0(x) \o {dt_2}}  \eq 2\om _1^{\pr}(x) \nonu\\
{d\om _1(x) \o {dt_2}} \eq 2\om _2^{\pr}(x) + (\om _0^2(x))^{\pr}
\lab{omega2}
\er
and
\br
{d\om _0(x) \o {dt_3}} \eq 3\om_2^{\pr}(x) + 6 \om _0 \om_0^{\pr} \nonu\\
{d\om _1(x) \o {dt_3}} \eq 3\om_3^{\pr}(x) + 6(\om _0(x)\om_1(x))^{\pr}
\lab{omega3}
\er
The first two equations \rf{omega2} are compatible with
\br
{dJ_k(x) \o {dt_2}} \eq - (J_k^2(x))^{\pr} - 2 \bar J^{\pr}(x)\nonu\\
{d\bar J_k(x) \o {dt_2}} \eq -2 (\bar J_k(x) J_k(x))^{\pr}
\lab{jota2}
\er
where
\be
\bar J(x) = \sum_{k=1}^{N} \bar J_k(x)
\lab{37}
\ee
Equations \rf{jota2} describe coupled multi-layered Benney equations studied by
Zakharov \ct{ZA}.
On the other hand the second set of equations \rf{omega3}, is compatible with
two sets of flows for $J_k$ and $\bj _k$, namely:
\br
{dJ_k(x) \o {dt_3}} \eq (J_k(x)^3)^{\pr} + 3(J_k(x)\sum_{l}\bj _l(x))^{\pr} +
3 \sum_{l}(\bj _l(x)J_l(x))^{\pr}\nonu\\
{d\bar J_k(x) \o {dt_3}} \eq 3(\bj _k(x)J_k^2(x))^{\pr} +
3\sum_l(\bj _l(x)\bj_k(x))^{\pr}
\lab{jota31}
\er
and
\br
{dJ_k(x) \o {dt_3}}\eq (J_k(x)^3)^{\pr} + 2(J_k(x)\sum_{l}\bj _l(x))^{\pr} +
4 \sum_{l}(\bj _l(x)J_l(x))^{\pr}\nonu\\
{d\bar J_k(x) \o {dt_3}} \eq 3(\bj _k(x)J_k^2(x))^{\pr} + 4\bj_k(x)
\sum_l(\bj_l(x))^{\pr} +2 (\bj _k(x))^{\pr}\sum_l \bj _l(x)
\lab{jota32}
\er
We can now define a first bracket structure in terms of $J_k$ and $\bj _k$ as
\br
\{\bar J_k(x),J_l(y)\}_{1} \eq -\d^{\prime}(x-y)\d_{kl}\nonu\\
\{ J(x)_k,J_l(y)\}_{1} \eq \{\bar J_k(x),\bar J_l(y)\}_{1} = 0
\er
Therefore \rf{35} generates under the first bracket the algebra
$\win1 \otimes \win1 \otimes \ldots$.

The corresponding Hamiltonian equations of motion:
\br
{dJ_k(x) \o {dt_r}} & =& \{J_k(x) \, , \, H_{r+1}\}_1 \nonu\\
{d\bj _k(x) \o {dt_r}} & =& \{\bj _k(x) \, , \, H_{r+1}\}_1
\lab{tr}
\er
reproduce for $r=2,3$  the flow equations  \rf{jota2} and \rf{jota31},
respectively.

 We can realize \rf{27l} through \rf{35} if we assume a second bracket
structure to be:
\br
\{\bar J_k(x),J_l(y)\}_2 &=& J_l(x)\d_{kl}\, \d^{\pr}(x-y) \nonu\\
\{\bar J_k(x),\bar J_l(y)\}_2 &=& 2 \bar J_l(x)\d_{kl}\, \d^{\pr}(x-y) +
\bar J_l^{\pr}(x)\d_{kl}\, \d (x-y)\nonu \\
\{J_k(x),J_l(y)\}_2 &=& 2\, \d^{\pr}(x-y)
\lab{33a}
\er
where the r.h.s. of the last equation is independent of $k$ and $l$.
We should point out at this stage that the above algebraic structure
violates the Jacobi identity.
However, it is satisfied at the level of $\om 's$ defined in
\rf{35}.
This second bracket can also be extended to the Hamiltonian framework
as follows:
\br
{dJ_k(x) \o {dt_r}}& =& \{J_k(x),H_{r}\}_2 \nonu\\
{d\bj _k(x) \o {dt_r}}& =&\{\bj _k(x),H_{r}\}_2
\lab{tr2}
\er
Taking $r=2,3$ we reproduce after using the algebra \rf{33a} the flow equations
\rf{jota2} and \rf{jota32}, respectively.
The system is therefore not bi-hamiltonian since the hierarchies of
equations \rf{jota2} relative to brackets 1 and 2 are different.

\subsection{The Current Algebra of WZNW Model}
The ordinary WZNW model associated to a Lie group $G$ possesses two
commuting chiral copies of the current algebra:
\be
\lcurl J_a(x) \, ,\, J_b(y) \rcurl = f_{ab}^c J_c(x) \d (x-y) +
k g_{ab} \, \d^{\pr}(x-y)
\lab{ordca}
\ee
where $f_{ab}^c$ are the structure constants of the Lie algebra $\lie$ of
$G$, and $g_{ab}$ is the Killing form of $\lie$.
The two chiral components of the energy momentum tensor
are of the Sugawara form
\be
T(x) = \sum_{a,b=1}^{dim G} g^{ab} J_a(x) J_b(x)
\ee
where $g^{ab}$ is the inverse of the Killing form. Such tensor
satisfies the Virasoro algebra with vanishing central term
\be
\lcurl T(x) \, ,\, T(y) \rcurl = 2 T(x)  \d^{\pr}(x-y) +  T^{\pr}(x)
\d (x-y)  \lab{virsuga}
\ee
The currents are spin one primary fields:
\be
\lcurl T(x) \, ,\, J_a(y) \rcurl = J_a (x)  \d^{\pr}(x-y)
\lab{primary}
\ee
Suppose now one has a self-commuting current ${\cal J}$,
$\lcurl {\cal J}(x) \, ,\, {\cal J}(y) \rcurl = 0$.
For the non compact WZNW model this current can, for instance, be the one
associated to a step operator $J(E_{\a})$ for any root $\a$ of $G$.
One then sees that the system
$(T,{\cal J})$ generates an algebra isomorphic to \rf{33} with the difference
 that now the last bracket relation is zero.
One can construct out of them the quantities  $w_n(x)
\equiv  T(x) {\cal J}^{n-2}$ satisfying the area preserving diffeomorphism
algebra, i.e.
\be
\{ w_n (x) \, , \, w_m (y) \} = \( n + m -2 \) w_{n+m-2} (x) \d^{\pr} (x - y)
+( m-1 ) \( w_{n+m-2 } (x)\)^{\pr} \d (x -y)  \phantom{......}
\lab{smallw}
\ee
By taking now a $U(1)$ subalgebra :
\be
\lcurl {\cal J}(x) \, ,\, {\cal J}(y) \rcurl = k g_{\scriptstyle {\cal J}}
\d^{\pr}(x-y) \ee
one sees that the $(T,{\cal J})$ system now generates an algebra which is
isomorphic to  \rf{33} where now the last bracket is different from zero.
The quantities $w_n$ introduced above will then generate a deformed
(nonlinear) area preserving diffeomorphism algebra, i.e.
\br
\lcurl w_{n}(x) \, ,\, w_{m} (y) \rcurl &=&
\( n + m -2 \) w_{n+m-2} (x) \d^{\pr} (x - y)
+( m-1 ) \( w_{n+m-2 } (x)\)^{\pr} \d (x -y) \nonumber\\
&-& k g_{\scriptstyle {\cal J}} (n-2)(m-2)w_{n-1}(x) \pa_x ( w_{m-1}(x)
\d (x-y))     \lab{smallwc}
\er

\sect{A Bose Construction of KP Hierarchy and Fa\'a di Bruno Polynomials}

In order to provide a two current realization of $\Win1$ and the
corresponding non-linear $\hWinf$ algebras we shall now give an
introduction to the \faa polynomials.
\subsection{Two-Boson KP Hierarchy and Fa\'a di Bruno Polynomials}
We now construct the KP hierarchy in terms of a pair of Bose currents
$J$ and $\bj$. We propose the following Lax components $W_n$:
\be
W_n(x) = (-1)^n \bj (x) P_n(J(x))
\lab{54}
\ee
given in terms of the \faa polynomials
\be
P_n(J) = (D + J)^n \cdot 1 = e^{-\Phi}\pa^n e^{\Phi}
\lab{55}
\ee
where $J(x) = \Phi^{\pr}(x)$ and $\pa^{n} = {\pa^n \o {{\pa x}^n}}$.
One easily recognizes in \rf{54} deformations of the area
preserving generators \rf{28}.

We associate to \rf{54} the Lax operator given by
\be
L=D+\bj {1\o{D+J}}=D+\sum_{n=0}^{\infty} W_n (x) \, D^{-1-n}
\lab{56}
\ee

{\sl \faa Polynomials and Their Properties.}

The technical analysis we are about to present relies rather
heavily on properties of \faa polynomials defined by
\be
P_n (J) \equiv e^{-\Phi} \pa^n e^{\Phi}
\lab{38}
\ee
It follows from \rf{38} that the generating functional for the
\faa polynomials is given by
\be
\exp \lcurl \sum_{k=1}^{\infty} \eps^k J^{(k-1)}/k! \rcurl
 = \sum_{k=0}^{\infty} P_k (J) \eps^k /k!
\lab{39}
\ee
which follows from
\be
e^{\Phi (x+\eps)- \Phi (x)}= \exp \lcurl \sum_{k=1}^{\infty} {\eps^k \o k!}
J^{(k-1)} \rcurl
\lab{40}
\ee
Consider now $f = \p^{\prime}_1$ , $g = \p^{\prime}_2$ and:
\br
\sum_{k=0}^n {n \choose k} P_k (f) P_{n-k} (g) &=& \sum_{k=0}^n {n \choose
k}e^{-\p_1}\pa^ne^{\p_1}e^{-\p_2}\pa^{n-k}e^{\p_2} \nonu\\
&=&
e^{-(\p_1+\p_2)}\sum_{k=0}^n {n \choose k}
\pa^ke^{\p_1}\pa^{n-k}e^{\p_2}\nonu\\
&=& e^{-(\p_1+\p_2)}\pa^ne^{\p_1+\p_2}
\er
where we had use Leibniz rule in the last step.  Therefore we find the identity
\be
P_n(f+g) = \sum_{k=0}^n {n \choose k} P_k (f) P_{n-k} (g), \qquad;\qquad
n=0,1,\ldots
\lab{41}
\ee
{}From \rf{38} it follows the recurrence relation
\be
\pa P_n = P_{n+1} - J P_n
\lab{42}
\ee
which suggests the alternative expression for the \faa polynomials:
\be
P_n(J) = (D+J)^n1
\lab{43}
\ee
satisfying \rf{42}.  The lowest order polynomials are
\be
P_0  = 1 \;\; ;\;\; P_1 = J \; \; ; \;\;
P_2 =  J^{2} +  J^{\pr}  \; \; ;\;\;
P_3 =    J^{3} + 3 J J^{\pr} +  J^{\pr \pr}  \;\;\; {\rm etc}
\lab{44}
\ee
Using Leibniz rule we also have
\be
\(D +J\)^n = e^{-\Phi} D^n e^{\Phi} = \sum_{l=0}^n {n \choose l}
e^{-\Phi} \pa^{n-l} e^{\Phi} D^l = \sum_{l=0}^n {n \choose l}
P_{n-l} (J) D^l
\lab{45}
\ee
Using equation \rf{4} we also have
\be
P_n (J) = e^{-\Phi} \pa^n e^{\Phi} = \sum_{l=0}^n (-1)^l {n \choose l}
e^{-\Phi} D^{n-l} e^{\Phi} D^l = \sum_{l=0}^n (-1)^l {n \choose l}
\(D +J\)^{n-l} D^l \lab{754}
\lab{46}
\ee

With the help of Leibniz rule we also obtain
\be
\(D-J\)^n = e^{\Phi} D^n e^{-\Phi} = \sum_{l=0}^n (-1)^{n-l} {n \choose l}
D^l \(\pa^{n-l} e^{\Phi}\) e^{-\Phi} = \sum_{l=0}^n (-1)^{n-l}
{n \choose l} D^l P_{n-l} (J) \lab{47}
\ee
Similarly we get
\br
\lefteqn{
P_n(J) \( D-J \)^m =\pa^n e^{\Phi} D^m e^{-\Phi}}\nonu\\
&=&
\sum_{l=0}^m (-1)^{m-l} {m\choose l} D^l e^{-\Phi} \pa^{m+n- l} e^{\Phi}
= \sum_{l=0}^m (-1)^{m-l} {m\choose l} D^l P_{m+n- l} (J)
\lab{48}
\er
from which we get
\be
P_n(J)\sum_{k=0}^m (-1)^k {m\choose k} D^k P_{m-k}(J) =
\sum_{k=0}^m (-1)^k {m\choose k} D^k P_{n+ m-k} (J)
\lab{49}
\ee
Again it follows from the Leibniz rule:
\br
D^n &=& D^n \( e^{\Phi} e^{-\Phi}\) =
\sum_{l=0}^n {n\choose l} e^{-\Phi} \( \pa^{n- l} e^{\Phi} \)
e^{\Phi} D^{l} e^{-\Phi} \nonu\\
&=& \sum_{l=0}^n {n\choose l} P_{l} (J) \( D - J \)^{n-l}
\lab{50}
\er
and using \rf{5} we find
\br
D^n &=&\( e^{-\Phi} e^{\Phi}\) D^n =
\sum_{l=0}^n (-1)^{l} {n\choose l} e^{-\Phi} D^{n- l} e^{\Phi} e^{-\Phi}
\pa^l e^{\Phi} \nonu\\
&=& \sum_{l=0}^n (-1)^{l} {n\choose l} \( D +J \)^{n-l} P_{l} (J)
\lab{51}
\er
{}From \rf{3} it follows
\be
e^{-\Phi}D^{-1} e^{\Phi} =
\sum_{l=0}^{\infty} (-1)^l e^{-\Phi} \pa^{l} e^{\Phi} D^{-l-1}=
\sum_{l=0}^{\infty} (-1)^l P_{l} (J) D^{-l-1}
\lab{52}
\ee
The r.h.s. is precisely $(D+J)^{-1}$. Therefore
\be
(D+J)^{-1} = e^{-\Phi}D^{-1}e^{\Phi}
\lab{53}
\ee
as expected from \rf{45}.

\subsection{The Algebraic Structure of Two-Boson KP Hierarchy}

Armed with the above technical details we proceed here to
calculate the algebra satisfied by the generators \rf{54}
in case when the canonical variables $J$ and $\bj$ satisfy the bracket
\rf{29}.
As a consequence of \rf{29} we clearly find:
\be
\{ \bj (x) \, , \, e^{\pm \Phi (y)} \} = \mp \d (x-y) e^{\pm \Phi (y)}
\lab{58}
\ee
The advantage of using the exponential representation \rf{38} is that it makes
relatively easy to calculate brackets between generators $W_n(x) = (-1)^n \bj
(x) e^{-\Phi}\pa^n e^{\Phi}$
\br
\{ W_n (x) \, , \, W_m (y) \}_1 &=& (-1)^{n+m} \bj (y) e^{-\Phi (y)} \pa_y^m
\( \d (x-y) \pa_y^n e^{\Phi (y)}\) \nonumber\\
&-& (-1)^{n+m} \bj (x) e^{-\Phi (x)} \pa_x^n \( \d (x-y) \pa_x^m e^{\Phi (x)}\)
\nonumber \\
&=& \sum_{k=0}^{m}(-1)^{n+m+k} {m\choose k} \bj (y) e^{-\Phi (y)} \pa_y^{m+n-k}
e^{\Phi (y)} \pa_x^k \d (x-y) \nonumber \\
&-& \sum_{k=0}^{n}(-1)^{n+m+k} {n\choose k} \bj (x) e^{-\Phi (x)} \pa_x^{m+n-k}
e^{\Phi (x)} \pa_y^k \d (x-y)
\lab{59}
\er
One recognizes in \rf{59} the first Gelfand Dickey structure written in the
Watanabe form \rf{15aa}

We will now show how to generate the second bracket structure from the
representation given by \rf{54}. This time the algebra of $J$ and $\bj$
is given by \rf{33}.
{}From \rf{33} it is possible to show that
\br
\lcurl P_n (x) \, ,\, P_m (y) \rcurl_2  = - c
\Bigg\lb \sum_{l=1}^n \sum_{p=1}^m (-1)^p {n \choose l} {m \choose p}
P_{n-l} (x) P_{m-p} (y) \pa^{l+p-1}_x \Bigg\rb \d (x-y)
\er
We obtain therefore the total second bracket for the generators in \rf{54} as
 the sums of linear and non linear terms
\br
\{ W_n (x) \, , \, W_m (y) \}_2 &=& \O^{(1)}_{nm} (W(x)) \d (x-y) \lab{62}\\
&-& c  \sum_{i=0}^{n-1}\sum_{j=0}^{m-1} (-1)^{n-1}{n \choose i} {m\choose j}
W_i (x) D_x^{n+m-i-j-1} W_j (x)  \d (x-y) \nonu
\er
for $n,m\geq 0$. The first term is the linear part which coincides with the
linear part of \rf{20}.

The result \rf{62} appears to be surprising since we recognize in the non
linear part of it only the Drinfeld-Sokolov structure from \rf{20} multiplied
by $c$, while the non linear part of the second Gelfand-Dickey bracket from
\rf{62} appears to be missing.
However one can show that \ct{2boson}:
\be
\{ W_n (x) \, , \, W_m (y) \}_2^{GD} \big\v_{\rm nonlinear} \,= \,
\{ W_n (x) \, , \, W_m (y) \}^{DS}
\lab{63}
\ee
That is, the non linear part of the second Gelfand-Dickey bracket is exactly
equal to the Drinfeld-Sokolov bracket. As a consequence of that we can rewrite
relation \rf{62} for the choice $c=2$ as
\be
\{ W_n (x) \, , \, W_m (y) \}_2 = \{ W_n (x) \, , \, W_m (y) \}_2^{GD} +
\{ W_N (x) \, , \, W_m (y) \}^{DS}
\ee
and therefore reproduce the form \rf{20}.

\subsection{Examples; Two loop WZNW, Conformal Toda Theories and
$W_{\infty} $ algebra}

Let us consider the WZNW model associated to a Kac-Moody
algebra whose conserved currents satisfy the two loop Kac-Moody algebra
\ct{AFGZ}:
\br
\lb J^m_a (x) \, , \, J^n_b (y) \rb &=& i f_{ab}^c J_c^{m+n}(x) \delta
(x - y)  + k g_{ab} \pa_x  \d (x - y)  \delta_{m,-n}
+ J^C (x) \d (x - y)  g_{ab} m \delta_{m,-n}
\nonu \\
\lb J^D (x) \, , \, J^m_a (y) \rb &=& m J^m_a (y) \d (x -y)
\nonu \\
\lb J^C (x) \, , \, J^D (y) \rb &=& k \pa_x  \d (x - y)
\lab{kma} \nonu\\
\lb J^C (x) \, , \, J^m_a (y) \rb &=& 0
\er
In this case the Sugawara energy-momentum tensor is
\be
T(x) = {1 \o {2}}\sum_{a,b=1}^{dim g}\sum_{n\eps Z}
g^{ab}J_a^n(x)J_b^{-n}(x) + J^D(x)J^C(x)
\ee
It is easy to show that $T(x)$ and $J^C(x)$ reproduce the algebra \rf{33}
with $h=c=0$  showing that the two-loop WZNW has the
$\winf$ structure.
In fact, this model possesses a bigger symmetry.
In order to show this let us consider the modified energy-momentum
tensor given as
\be
U(x) = T(x) +{\pa \o {\pa x}}(2J_{\hat \d }(x) + h J^D(x) +{2\hat \d ^2 \o
{h}}J^C(x))
\lab{mod}
\ee
where $J_{\hat \d}(x) = kTr(\hat g^{-1}(x)\pa \hat g(x) \hat \d .H)$ , $\hat
\d = {1 \o {2}}\sum_{\a >0}{\a \o {\a ^2}}$ and $\hat g$ denotes a group
element generated by exponentiation of the Kac-Moody algebra (see \ct{AFGZ}
for further details), $h$ is the Coxeter number of the underlying Lie
algebra.
It can be shown that the operators $U(x)$ given in \rf{mod} and $J^C(x)$
satisfy precisely the structure \rf{33} with $c=0$ and the
deformation parameter $h$ being the Coxeter number.  In this case we have
$\Winf$ algebra as the symmetry of the problem \ct{2boson}.  In this last
example it is known that the modification of the energy-momentum tensor allows
the Hamiltonian reduction of the two-loop WZNW model to the Conformal Affine
Toda model.  This model, inherits in turn the same symmetry structure
$\Winf$ in contrast to the usual Conformal Toda models which has lost
invariance under $\winf$.

\sect{``Gauge" Equivalence between various KP Hierarchies}
The fundamental result of \ct{ANPV} was the proof
that all three hierarchies labelled by $\ell=0,1,2$
are ``gauge" equivalent via
generalized Miura transformations.
Here we focus on the link between two \KPo and \mKPa systems discussed above.
Reference \ct{ANPV} presented a gauge transformation
between those two systems, which  mapped the underlying first bracket
structures into each other.
Explicitly this map is provided by $G$ from \rf{gaug1}
where it mapped $L \stackrel{G}{\longrightarrow} K$ with $K$ being in the
KP hierarchy.
We have seen that this map connects various KP flow equations.
In \ct{ANPV} it was shown that the map transforms two different
first bracket structures into each other.
Take namely vectors $Q,V$ from \rf{7} and notice that only
$Q_{\ge 0}, V_{\ge 0}$ contribute to $\me{K}{Q}$ and $\me{K}{V}$.
Proof given in \ct{ANPV} verified the validity of the following identity:
\be
\langle K\mid \lb Q_{\ge 0}\, ,\, V_{\ge 0} \rb \rangle
=
\{ \langle K \mid Q \rangle \, , \, \langle K \mid V \rangle \}_1^{\ell=0}
=
\{ \langle G^{-1} L G
\mid Q \rangle \, , \, \langle G^{-1} L G \mid V\rangle \}_1^{\ell=1}
\lab{gaug2}
\ee
where in the last expression the bracket was understood according to
definition in \mKPa \ct{ANPV}.
The proof is based on the definition of the first bracket structure as a
Lie-Poisson $R$-bracket
for functions $F, H \in C^{\infty} \(\dlie, \IR \)$:
\be
\{ F \, , \, H \}_R (L) =
\me{L}{{\sbr{\nabla F (L)}{\nabla H (L)}}_R}   \lab{Rbra}
\ee
where the gradient $\nabla F\,: \, \dlie \ra \lie$ is defined by the
standard formula  given in \ct{AKS,ANPV}.

It is in this sense \KPo and \mKPa are called ``gauge" equivalent.

As we will see
the mapping \rf{gaug1} transforming the Poisson bracket structure of
\KP into that
of \mKPa and vice versa
deserves a name of the
generalized Miura transformation.

As the first application
let us connect the two-boson KP hierarchy Lax operator $L_{J}^{(1)}$
with the Lax operator \rf{56} expressed in terms of \faa polynomials.
Consider namely the gauge transformation between \mKPa and \KPo
generated by $\P$ such that $\P^{\pr} = J$:
\be
L_{J} = e^{-\P} L_{J}^{(1)} e^{\P} = D + \bj \(D + J\)^{-1} =
D + \sumi{n=0} (-1)^n \bj P_n (J) D^{-1-n}
 \lab{faalax}
\ee
where $P_n (J)= (D + J)^n \cdot 1$ are as before the \faa polynomials.
As a corollary of the symplectic character of the ``gauge"
transformation used in \rf{faalax}, we conclude that
$u_n = (-1)^n \bj P_n (J) $ belonging to the Lax operator of \KPo
must satisfy the Poisson-bracket $\Win1\,$ algebra described by the
form $\O^{(0)}$ according to \rf{15aa} \ct{BAK85,2boson}.
This represents an elegant version of the technical proof given in section 4.
It is possible to introduce a deformation parameter into
the \faa representation of $\Win1\,$ algebra by redefining $u_n$ to
$u_n (h) = (-1)^n \bj (h D + J)^n \cdot 1$ \ct{2boson}.
Now the semiclassical limit is simply obtained by taking $h \to 0$
in $u_n (h)$ and yields the generators of ${\bf w_{1+\infty}}$
algebra.

We now will discuss gauge equivalence between various two-boson hierarchies.

\subsection{Gauge Equivalence of Various Two-Boson KP Hierarchies.}

\lskip
{\sl The Non-Linear Schroedinger Hierarchy.}
The non-linear Schroedinger (NLS) system is a constrained KP system
described by:
\be
L_{NLS}\, =\, D + {\bar \psi} D^{-1} \psi
\lab{akns}
\ee
with the following (non-trivial) flow equations at the lowest level:
\be
{d \o d t_2} { {\bar \psi} \choose \psi} = \( \begin{array}{c}
{\bar \psi}^{\pr\, \pr} + 2 \psi {\bar \psi}^2 \\
-\psi^{\pr\, \pr} - 2 {\bar \psi} \psi^2
\end{array} \) \quad ; \quad
{d \o d t_3} { {\bar \psi} \choose \psi} = \( \begin{array}{c}
{\bar \psi}^{\pr\, \pr\, \pr } + 6 \psi {\bar \psi} {\bar \psi}^{\pr}  \\
\psi^{\pr\,\pr\, \pr} + 6  {\bar \psi} \psi \psi^{\pr}
\end{array} \)           \lab{akns1}
\ee
These first equations of the NLS hierarchy can be reproduced by the
Hamiltonian approach. The first Hamiltonians functions obtained via usual
definitions are
\be
H_1 = \int {\bar \psi} \psi \quad ; \quad H_2 = \int {\bar \psi}^{\pr}
\psi \quad;\quad H_3 = \int \( {\bar \psi}^{2} \psi^2 -
{\bar \psi}^{\pr}
\psi^{\pr} \)
\lab{akns2}
\ee
while the first two bracket structures are given by:
\br
\{\bar \psi (x),\psi (y)\}_1 &=& \d (x-y) \nonu\\
\{\bar \psi (x),\bar \psi (y)\}_1 &=& \{ \psi (x),\psi (y)\}_1 =0
\lab{69}
\er
and
\br
\{ \psi (x) \, , \, \psi (y) \}_2 &=& -2
\psi (x) \pa^{-1}_x \psi (x) \d (x-y)\nonu\\
\{ \psi (x) \, , \, {\bar \psi} (y) \}_2 &=& \d^{\pr} (x-y) +2 \psi (x)
 \pa^{-1}_x {\bar \psi} (x) \d (x-y)
\lab{72}\\
\{ {\bar \psi} (x) \, , \, {\bar \psi} (y) \}_2 &=& -2 {\bar \psi} (x)
 \pa^{-1}_x {\bar \psi} (x) \d (x-y)\nonu
\er
The last structure can also be rewritten as:
\br
\{ \psi (x) \, , \, \psi (y) \}_2 &=& - \psi (x)  \psi (y) \epsilon
(x-y)\nonu\\
\{ \psi (x) \, , \, {\bar \psi} (y) \}_2 &=& \d^{\pr} (x-y) + \psi (x) {\bar
\psi} (y) \epsilon (x-y)
\lab{73}\\
\{ {\bar \psi} (x) \, , \, {\bar \psi} (y) \}_2 &=& - {\bar \psi} (x) {\bar
\psi} (y) \epsilon (x-y)\nonu
\er
where $\epsilon (x-y)$ is the sign function. It is easy to see that the
formulas \rf{72} and \rf{73} satisfy the Jacobi identity.
One can show that NLS hierarchy for independent
${\bar \psi}$ and $\psi$ is equivalent to \faa two-boson KP.
The proof is based, in the spirit of \ct{ANPV}, on
establishing gauge transformation between two hierarchies (see also
\ct{oevels,bonoxiong,west,schiff}).
We show now the argument to illustrate the power of gauge
transformation argument in the KP setting.
Consider
\be
L_{NLS} \to G^{-1} L_{NLS} G = G^{-1} D G +  G^{-1} {\bar \psi} \,
D^{-1} \psi G
\lab{akns3}
\ee
Take $ G^{-1} = \psi$, which leads to:
\be
G^{-1} L_{NLS} G = \psi \, D \psi^{-1} \, +  {\bar \psi} \, \psi \, D^{-1}
= D + \psi \, ( \psi^{-1} \, )^{\pr} \, +  {\bar \psi}\, \psi \, D^{-1}
\lab{akns4}
\ee
Clearly the gauge transformed $L_{NLS}$ is an element of ${\rm KP}_1$
hierarchy and it is therefore natural to introduce now new variables such
that
\br
\bj (x) &=&\bar \psi (x) \psi(x) \nonu\\
J(x) &=& {\psi ^{\prime}(x) \o {\psi (x)}}
\lab{68}
\er
and the inverse relation being $\psi = \exp \(\int J\)$ and
${\bar \psi} = \bj \exp -\(\int J\)$.
Since we now have established a gauge equivalence between two hierarchies
it is clear that the first bracket structure in \rf{69}
leads to \rf{29} with the generators of a linear
$\Win1$ algebra being
\be
W_n (x) = (-1)^n \bar \psi (x) \psi ^{(n)}(x)
\lab{71}
\ee
where $\psi^{(n)}(x) = {d^n\psi \o {{dx}^n}}$, while
\be
P_n(J(x)) = {\psi ^{(n)}(x) \o {\psi (x)}}
\lab{70}
\ee

The second bracket structure in \rf{72} leads to
\rf{33} with $c=2, h=1$ and correspondingly non-linear $\hWinf$.
If we only took the linear structure in \rf{72}
(i.e. $\pbr{{\bar \psi}\, (x)}{\psi\,(y)} = \d^{\pr} (x-y)$) we would have
induced \rf{33} in its ``un-deformed" form  with the last equation of
\rf{33} being zero, corresponding to $\O^{(1)}$ or
$\Winf$.

We can also generalize above construction by adding N independent copies of
 currents:
\be
W_n (x) = (-1)^{n}\sum_{k=1}^{N} \bj _k(x) P_n(J_k(x)) = (-1)^{n}\sum_{k=1}^N
\bar \psi_k(x)\psi_k^{(n)}(x)
\lab{74}
\ee
The first bracket given by
\br
\{\bar \psi_k (x),\psi_l (y)\}_1 &=& \d_{kl}\d (x-y) \nonu\\
\{\bar \psi_k (x),\bar \psi_l (y)\}_1 &=& \{ \psi_k (x),\psi_l (y)\}_1 =0
\lab{75}
\er
The second bracket structure generating the non linear $\hWinf$ can
 be obtained through the algebra:
\br
\{ \psi_k (x) \, , \, \psi_l (y) \}_2 &=& - \psi_k (x)  \psi_l (y) \epsilon
 (x-y)\nonu\\
\{ \psi_k (x) \, , \, {\bar \psi}_l (y) \}_2 &=& \d_{kl}\d^{\pr} (x-y) + \psi_k
 (x) {\bar \psi}_l (y) \epsilon (x-y)
\lab{76}\\
\{ {\bar \psi}_k (x) \, , \, {\bar \psi}_l (y) \}_2 &=& - {\bar \psi}_k (x)
 {\bar \psi}_l (y) \epsilon (x-y)\nonu
\er
and through the definition
\br
J_k (x) \equiv {\psi^{\pr}_k (x) \o \psi_k (x)} \, \, \, , \, \, \,
\bj_k (x) \equiv {\bar \psi}_k(x)\psi_k (x)
\lab{77}
\er
which reproduce the algebra \rf{33} with an anomaly term in the first relation,
 i.e.
\br
\{\bar J_k(x),J_l(y)\}_2 &=& J_l(x)\d_{kl}\d^{\prime}(x-y)  -
\d_{kl}\d^{\pr\pr} (x-y)\nonu\\
\{\bar J_k(x),\bar J_l(y)\}_2 &=& 2 \bar J_l(x)\d_{kl}\d^{\prime}(x-y) +
\bar J_l^{\prime}(x)\d_{kl}\d (x-y)\nonu \\
\{J_k(x),J_l(y)\}_2 &=& 2\d^{\prime}(x-y)
\lab{78}
\er
As in the discussion of the multi-layered Benney equations the above
algebra does not satisfy the Jacobi identity.
Correspondingly the brackets for the fields $\psi$'s and ${\bar \psi}$'s
given in \rf{76} will violate Jacobi as well.

The brackets \rf{75} and \rf{76} lead to a system of coupled non-linear
Schroedinger equations \ct{svi}:
\br
{d\psi_m(x) \o {dt}} &=& \{\psi_m(x),H_3\}_1 =\{\psi_m(x),H_2\}_2 =
-\psi^{\pr \pr}_m(x) - 2 \psi_m(x)\sum_{k=1}^{N}{\bar \psi}_k(x)\psi_k(x)
\nonu\\
{d{\bar \psi}_m(x) \o {dt}} &=& \{{\bar \psi}_m(x),H_3\}_1 =
\{{\bar \psi}_m(x),H_2\}_2 = {\bar \psi}^{\pr \pr}_m(x) +
2{\bar \psi}_m(x)\sum_{k=1}^{N}{\bar \psi}_k(x)\psi_k(x)
\er
The hierarchies of these equations relative to brackets 1 and 2 are
different.

\subsection{Quadratic Two-Boson KP Hierarchy and Generalized Miura
Transformation}
\lskip
{\sl Quadratic Two-Boson KP Hierarchy.}
Here we call quadratic two-boson KP hierarchy
the construction presented by Wu and Yu \ct{wuyu,didier}
in order to realize $\hWinf$ as a hidden current algebra in the
2d $SL(2, \IR )/U(1)$ coset model.
Construction is based on the pseudo-differential operator:
\be
L_{\sj} = D + \bsj\, \(D - \sj \,- \bsj \,\)^{-1} \sj
\lab{sjlax}
\ee
Let us discuss the Hamiltonian structure first.
The three lowest Hamiltonian functions are:
\br
H_{\sj\;1} \eq \int \sj\, \bsj \; dx \lab{hsj1} \\
H_{\sj \;2} \eq \int \(- \sj^{\pr}\, \bsj + \sj^{2}\, \bsj +\sj\, \bsj^{2} \,
\) \, dx \lab{hsj2} \\
H_{\sj\;3} \eq \int \( \sj^{\pr \pr}\, \bsj -3 \sj \,\sj^{\pr}\, \bsj
- 2 \sj^{\pr}\, \bsj^2 - \sj\, \bsj\, \bsj^{\,\pr} + \sj^{3} \, \bsj
+ 3 \sj^{2} \, \bsj^{\,2} + \sj\, \bsj^{\,3} \, \) \, dx
 \lab{hsj3}
\er
Among the Hamiltonian structures only second and third are local and are
given by
\br
\{ \sj \, (x) \,, \, \bsj\, (y) \}_2 \eq \d^{\pr} (x-y)  \lab{p1sj} \\
\{ \sj \, (x) \,, \, \sj\, (y) \}_2 \eq \{ \bsj \,(x) \,, \, \bsj\, (y) \}_2
= 0 \nonu
\er
and
\br
\{ \sj \, (x) \,, \, \bsj\, (y) \}_3 \eq \( \bsj \, (x) + \sj\, (x) \)
\d^{\pr} (x-y) + \sj^{\, \pr} (x) \d (x-y) - \d^{\pr\pr} (x-y)  \lab{p2sj} \\
\{ \sj \, (x) \,, \, \sj \, (y) \}_3 \eq 2 \sj \, (x) \d^{\pr} (x-y) +
\sj^{\, \pr} (x) \d (x-y) \nonu\\
\{ \bsj \, (x) \,, \, \bsj \, (y) \}_3 \eq 2 \bsj \, (x) \d^{\pr} (x-y) +
\bsj^{\, \pr} (x) \d (x-y)\nonu
\er
\prop  The Hamiltonian structure corresponding to the Lax operator
$L_{\sj}$ in \rf{sjlax} is invariant under the following
two transformations:
\br
\sj & \to & \bsj - {\sj^{\pr} \o \sj} \qquad \quad \quad \bsj \to \sj
\lab{sjtransf} \\
\bsj &\to & \sj + { \bsj^{\,\pr} \o \bsj \, }\qquad \quad\quad \sj \to \bsj
\lab{sjtransf2}
\er
\proof One verifies relatively easily that both bracket structures
\be
P_{\sj\;2}= \left(\begin{array}{cc}
0 & D \\
D & \; 0 \end{array}
\right) \;\; , \;\;
P_{\sj \;3}=\left(\begin{array}{cc}
D \sj \,+ \sj\, D & \; -D^2 + D \sj\, + \bsj \, D \\
D^2 + \sj\, D+ D \bsj \, &\; D\bsj\, + \bsj \, D \end{array}
\right)
\lab{p1sjp2sj}
\ee
corresponding to \rf{p1sj} and \rf{p2sj} are invariant under the
transformations \rf{sjtransf} and \rf{sjtransf2}.
Since $P_1= P_2 P_3^{-1} P_2$, a recurrence matrix $P_2 (P_1)^{-1}$ and
all remaining higher hamiltonian structures must therefore remain invariant
under \rf{sjtransf} and \rf{sjtransf2}.
This in principle completes the proof. One can also verify directly that
all three Hamiltonians \rf{hsj1}, \rf{hsj2}, \rf{hsj3} are invariant under
\rf{sjtransf} and \rf{sjtransf2} too.
Hence we conclude that the Lax operators given by
\be
L_{\sj} = D + \sj\, \(D - \sj \,- \bsj \,+ {\sj^{\,\pr} \o \sj} \)^{-1}
\( \bsj - {\sj^{\pr} \o \sj} \)
\lab{sjlax1}
\ee
and
\be
L_{\sj} = D + \( \sj + {\bsj^{\, \pr} \o \bsj} \)
\(D - \sj \,- \bsj \,-{\bsj^{\,\pr} \o \bsj} \)^{-1}  \, \bsj
\lab{sjlax2}
\ee
lead to the same Hamiltonian functions as \rf{sjlax}.

\lskip
{\sl Gauge Equivalence between \faa and Quadratic Two-Boson Hierarchies.
Generalized Miura Map.}

We apply on $L_{\sj}$ from \rf{sjlax} the gauge transformation
$\exp \( \p +\bp \)$ with result
\be
L_{\sj} \to \exp (- \p - \bp )\,L_{\sj}\, \exp (\p +\bp ) =
D + \sj \,+ \bsj\, + \bsj \, D^{-1} \sj
\lab{gaugea}
\ee
which is already an object in ${\rm KP}_1$ hierarchy.
Acting furthermore with the gauge transformation $\exp \(- \ln \sj\, \)$
we obtain finally the object in the \faa hierarchy.
\br
L_{\sj} \to \exp \( \ln \sj\, \) L_{\sj} \exp \(- \ln \sj\, \)
= D + \sj \,+\bsj \, + \sj \, ( \sj^{-1} \, )^{\pr} \, +  \bsj \, \sj \,
D^{-1} = D - J + \bj D^{-1}   \lab{lsjgauge}
\er
where we have introduced
\br
J \eq - \sj \,- \bsj \, + {\sj^{\,\pr}  \o \sj}  \nonu \\
\bj \eq \bsj \, \sj            \lab{gmiura}
\er
One can now verify explicitly that with the bracket structure given
\rf{p2sj} variables defined in \rf{gmiura} satisfy
the second bracket structure of \faa hierarchy \rf{33} with $c=2$ and
$h=1$ leading as shown in \ct{2boson} to $\hWinf$.
This is a short proof for the quadratic two-boson KP hierarchy \ct{wuyu}
system realizing $\hWinf$.
We now have obtained a {\bf Miura} transform for two-bose hierarchies in
form of \rf{gmiura} which generalizes the usual Miura transformation
between the one-bose KdV and mKdV structures (as shown below).

Of course the higher hamiltonian structures of quadratic two-boson hierarchy
are being mapped by \rf{gmiura} to their counterparts in \faa hierarchy.
This is also true for the Hamiltonian functions as one can see comparing
\rf{hJ} to \rf{hsj1}, \rf{hsj2}, \rf{hsj3}.

Let us go back to the alternative expression \rf{sjlax1} for the quadratic
two-boson hierarchy. It can be rewritten under multiplication by
 $1=\sj\,\sj^{\,-1}$ from the right and left as follows
\be
L_{\sj} = 1\, L_{\sj}\, 1 = \sj\, \sj^{\,-1}\,L_{\sj}\, \sj\, \sj^{\,-1}\,
= D + \, \(D - \sj \,- \bsj \,\)^{-1} \( \bsj\, \sj \,- \sj^{\,\pr} \, \)
\lab{sjlaxa}
\ee
Next step is to gauge transform \rf{sjlaxa} from KP to ${\rm KP}_1$ hierarchy
by acting with gauge transformation generated by $\exp ( \p + \bp )$ obtaining
\be
L_{\sj} \sim \exp \(- \p - \bp \) \,L_{\sj} \exp \( \p + \bp \)
= D + \sj \,+\bsj \,+\, D^{-1} \( \bsj\, \sj \,- \sj^{\,\pr} \, \)
\lab{sjlaxb}
\ee
which is of the form of the \faa hierarchy (up to conjugation) with
$ J =- \sj \,-\bsj \,$ and $ \bj = \bsj\, \sj \,- \sj^{\,\pr} \, $,
which is equal to what was done in \ct{2boson} chapter 3.3.
Note that under \rf{sjtransf2} this is transformed into
$ J = -\sj \,-\,\bsj \, - {\bsj{\,\pr} \o \bsj }\,$ and
$ \bj = \bsj\, \sj \,$ differing from \rf{gmiura} by a conjugation
$\sj\, \leftrightarrow \bsj \,$.

Similarly for \rf{sjlax2} we find
\be
L_{\sj} = \bsj^{\,-1} \bsj \, L_{\sj} \bsj^{\,-1} \bsj \,=\,
D + \(\bsj\, \sj + \bsj^{\, \pr} \,\)
\(D - \sj \,- \bsj \,\)^{-1}
\lab{sjlaxc}
\ee
The same transformation as in \rf{sjlaxb} gives
\be
L_{\sj} \sim \exp \(- \p - \bp \) \,L_{\sj} \exp \( \p + \bp \)
= D + \sj \,+\bsj \,+\, \(\bsj\, \sj + \bsj^{\, \pr} \,\) D^{-1}
\lab{sjlaxd}
\ee
producing ${\rm KP}_1$ object with
$ J =- \sj \,-\bsj \,$ and $ \bj = \bsj\, \sj \,+\bsj^{\,\pr} \, $
This time under \rf{sjtransf} these variables are transformed into
$ J = -\sj \,-\,\bsj \,+ {\sj{\,\pr} \o \sj }\,$ and
$ \bj = \bsj\, \sj \,$ being precisely a transformation from \rf{gmiura}.

We see that because of \rf{sjtransf} and \rf{sjtransf2} there is an ambiguity
in the possible form of generalized Miura transformation and \rf{gmiura} can
appear also in other forms all of them connecting the Poisson bracket
structure of \faa hierarchy with the corresponding Poisson bracket structure
of the quadratic two-boson hierarchy.

\lskip
{\sl Quadratic KP Hierarchy and the NLS Systems.}
We note first that the NLS system is also gauge equivalent to quadratic
KP hierarchy if we make in \rf{akns4} a substitution ${\bar \psi} =
\bsj\, \exp (\phi +\bp)$  and $ \psi = \exp (-\phi- \bp) \sj$ or inversely
$\psi^{\pr}/\psi = - \sj\, - \bsj \, + \sj^{\, \pr}/\sj$
and ${\bar \psi} \psi = \bsj\, \sj$.
This relation takes the following simple form in terms of the gauge
transformation acting on the Lax operator
$L_{\sj} =D + \sj \,+ \bsj\, + \bsj \, D^{-1} \sj$:
\be
e^{(\p + \bp)} \(D + \sj \,+ \bsj\, + \bsj \, D^{-1} \sj \)
e^{-(\p + \bp)}= D + {\bar \psi} D^{-1} \psi
\lab{sjtonls}
\ee
Changing the gauge function from $\exp{(\p + \bp)}$ to
$\exp{(-\p + \bp)}$ we establish link with so-called derivative Non-Linear
Schroedinger (dNLS) system:
\br
e^{(-\p + \bp)} \(D + \sj \,+ \bsj\, + \bsj \, D^{-1} \sj \)
e^{(\p -\bp)}\eq
D + 2 \sj\, + \bsj\, e^{(-\p + \bp)} D^{-1} \sj\, e^{(\p -\bp)} \nonu\\
\eq D + 2 r q + \( r q^2 + q^{\pr}\) D^{-1} r
\lab{sjtodnls}
\er
where we have introduced new variables \ct{schiff}:
\br
j(x) &=& q(x)r(x) \nonu\\
\bar j(x) &=& q(x)r(x) +{q^{\pr}(x) \o {q(x)}}= j(x) +{q^{\pr}(x) \o {q(x)}}
\lab{j}
\er
The Hamiltonian $H_{\sj\;1}$ is written in terms of these variables as
\be
H_{\sj\;1} = \int (q^2(y)r^2(y) + q^{\pr}(y)r(y))dy
\lab{h1}
\ee
We now propose the following algebraic structure
\br
\{q(x),r(y)\}_3 &=& \d ^{\pr}(x-y) \nonu\\
\{q(x),q(y)\}_3 &=&  \{r(x),r(y)\}_3 = 0
\lab{br3}
\er
Using the definitions \rf{j} we find exactly the third bracket structure
given in eq. \rf{p2sj}.

The corresponding equations of motion
\br
\dot q(x) = \{q(x), H_1\}_3 &=& q^{\pr \pr}(x) + 2(q^2(x)r(x))^{\pr} \nonu\\
\dot r(x) = \{r(x), H_1\}_3 &=& -r^{\pr \pr}(x) + 2(r^2(x)q(x))^{\pr}
\lab{eqmotion}
\er
correspond to the derivative NLS equations described in \ct{KauNew}.

Define now, according to Kaup and Newell \ct{KauNew}
\br
R(x) &=& q(x)e^{\mu (x)} \nonu\\
Q(x) &=& r(x)e^{-\mu(x)} \nonu\\
\mu (x) &=&\int_{-\infty}^{x} q(y)r(y)dy
\lab{RQ}
\er
It then follows that
\br
H_{\sj\;1} &=& \int R^{\pr}Q \nonu\\
H_{\sj\;2} &=& -\int (R^{\pr}Q^{\pr} - R^{\pr}RQ^2)
\er
which coincides with expressions (42b) and
(42c) of \ct{KauNew} up to factors $i$.

The second  bracket \rf{p2sj} can be realized by
\br
\{q(x),q(y)\}_2 &=& q(x)q(y)\epsilon (x-y) \nonu\\
\{r(x),q(y)\}_2 &=&  -\d (x-y) - r(x)q(y)\epsilon (x-y)\nonu \\
\{q(x),r(y)\}_2 &=& \d (x-y) - r(y)q(x)\epsilon (x-y)\nonu\\
\{r(x),r(y)\}_2 &=& r(x)r(y)\epsilon (x-y)
\lab{br2}
\er
The equations $ \dot q(x) = \{q(x)\, ,\, H_{\sj\;2}\}_2 $ and
$\dot r(x) = \{r(x) \,, \, H_{\sj\;2} \}_2$
reproduce \rf{eqmotion} showing that the system is indeed bi-hamiltonian.

\sect{Reduction to One-boson KdV Systems}
We here apply the Dirac reduction scheme to obtain one-boson hierarchies
from two-boson hierarchies. The general feature will be a transformation
of some two-boson Hamiltonian equations of motion
\be
\partder{{\cal O}}{t_r} = \{ {\cal O} \,, \, H_r \}_2
\lab{2bosham}
\ee
(where ${\cal O}$ denote original degrees of freedom)
to one-boson Hamiltonian system according to the Dirac scheme:
\be
\partder{X}{t_r} = \{ X \,, \, H^{D}_r \}_{Dirac}
\lab{diraham}
\ee
with $X$ denoting a surviving one-boson degree of freedom.
Another general feature would be that reduction would take the Lax operator
from ${\rm KP}_1$ hierarchy to the conventional ${\rm KP} $ hierarchy.
\lskip
{\sl KdV Hierarchy.}
Consider the Dirac constraint: $\Theta = J=0$ for system in \rf{faalax}.
First let us discuss the resulting Dirac bracket structure.
We find:
\be
\{ \bj (x) \, , \, \bj (y) \}_2^{D} = \{ \bj (x) \, , \, \bj (y) \}_2
- \int dz dz^{\pr}\{ \bj (x) \, , \, \Theta (z) \}_2
\{ \Theta (z) , \Theta (z^{\pr} )
\}_2^{-1} \{ \Theta (z^{\pr}) \, , \, \bj (y) \}_2
\lab{dira1}
\ee
which yields
\be
\{ \bj (x) \, , \, \bj (y) \}_2^{D} =
2 \bj (x) \d^{\pr} (x-y) +\bj^{\pr} (x) \d (x-y) + \h \d^{\pr \pr \pr} (x-y)
\lab{dira2}
\ee
The reduced Lax operator looks now as:
\be
l_{J} = D + \bj D^{-1}  \lab{faa3}
\ee
and the corresponding (non-zero) lowest Hamiltonian functions $H^{KdV}_r
\equiv \Tr l_{J}^r /r$ are
\be
H^{KdV}_{1} = \int \bj \, dx \quad ;\quad H^{KdV}_{3} = \int \bj^2 \, dx
\quad ;\quad H^{KdV}_{5} = \int \( 2 \bj^3 + \bj \bj^{\pr \pr} \)\, dx
\lab{kdvham}
\ee
Moreover one checks that the flow equation:
\be
{\d l_J}/ {\d t_r} = \sbr{(l_J^r)_{+} }{l_J}
\lab{flowkdv}
\ee
gives
\be
{\d \bj}/ {\d t_1}= \bj^{\pr} \qquad;\qquad {\d \bj}/ {\d t_3} =
\bj^{\pr \pr \pr} + 6 \bj \bj^{\pr}
\lab{kdveqs}
\ee
where the second equation reproduces the famous KdV equation.
This equation can also be obtained by inserting $X=\bj$ and $H^{KdV}_{3}$
into \rf{diraham}.
\lskip
{\sl mKdV Hierarchy.}
Now consider the quadratic two-boson hierarchy with Lax given in \rf{sjlax},
\rf{sjlax1} or \rf{lsjgauge}.
We choose as a Dirac constraint: $\theta = \sj \,+ \bsj\, = 0$.
The resulting Dirac bracket structure is:
\be
\{ \sj \, (x) \, , \, \sj \, (y) \}_2^{D} =
- \int dz dz^{\pr} \{ \sj \, (x) \, , \, \theta (z) \}_2
\{ \theta (z) , \theta (z^{\pr} )
\}_2^{-1} \{ \theta (z^{\pr}) \, , \, \sj \, (y) \}_2 = - \h \, \d^{\pr} (x-y)
\lab{dira3}
\ee
and the reduced Lax operator is:
\be
l_{\sj} = D - \sj \, D^{-1} \sj \, = D + \sumi{n=0} (-1)^{n+1} \sj\,
\sj^{\,(n)} D^{-1-n}
 \lab{mkdvlax}
\ee
Note that imposing the constraint $\theta=0$ on the equivalent Lax
operators from \rf{sjlax1} and \rf{sjlax2} respectively, we get:
\br
l_{\sj} \eq L_{\sj} \bv_{\theta=0} = D + \sj \(D \,+\,
{\sj^{\,\pr} \o \sj} \)^{-1}  \, \( - \sj\, - \,{\sj^{\, \pr} \o \sj} \)
= D + D^{-1} \( - \sj^{\, 2} - \, \sj^{\, \pr}\)
\lab{sjlaxreda} \\
l_{\sj} \eq L_{\sj} \bv_{\theta=0} = D -\( \sj\, +\,{\sj^{\, \pr} \o \sj} \)
\(D \,-\, {\sj^{\,\pr} \o \sj} \)^{-1}  \, \sj \,
= D +  \( - \sj^{\, 2} - \, \sj^{\, \pr}\) \, D^{-1}
\lab{sjlaxredb}
\er
The last equalities in \rf{sjlaxreda} and \rf{sjlaxredb}
were obtained using the trick of multiplying $l_{\sj}$
by $1= \sj\, \sj^{\, -1} =  \sj^{\, -1} \,  \sj\,$ from both sides.
Obviously we could express everywhere $\sj\,$ by $- \bsj\,$ hence
the one-boson system must be invariant under transformation
$\sj \, \leftrightarrow - \sj \,$.
The flow equations calculated as in \rf{flowkdv} are
\be
{d \sj\, \o d t_1} = \sj\,^{\pr} \quad;\quad {d \sj\, \o d t_2} = 0
\quad ;\quad
{d \sj\, \o d t_3} = \sj\,^{\pr \pr \pr}  + 6 \sj\,^2(\sj\,)^{\pr}
\lab{1bosflow}
\ee
Hence the flow equation for $ d \sj / d t_3$ is the mKdV equation.
Furthermore the mKdV equation could also be obtained from Hamiltonian
$H_3$ defined in a standard way:
\br
H_{1}^{mKdV} \eq - \int \sj\,^2 \, dx \quad ;\quad
H_{3}^{mKdV} = \int \(\sj\,^4 - \sj\,\sj^{\, \pr\pr}\)\, dx \nonu \\
H_{5}^{mKdV} \eq - \int \( 2 \sj\,^6 + 10
\sj\,^2 (\sj\,^{\pr})^2 + \sj\, \sj\,^{(IV)} \) \, dx
\lab{mkdvham}
\er
(and zero for even indices).
Because of existence of symmetry described in \rf{sjtransf} (and
\rf{sjtransf2}) we could equivalently impose the constraints
$\theta_1 = \sj \,+ \bsj\,- \sj^{\, \pr} /\sj\,  = 0$ or alternatively
$\theta_2 = \sj \,+ \bsj\,+\bsj^{\, \pr} /\bsj\, = 0$
without changing the Dirac bracket structure and the constraint manifold.
Imposing $\theta_1= 0 $ on the Lax operator in \rf{sjlax} we get
\be
l_{\sj} = D + \( - \sj\, + {\sj^{\, \pr} \o \sj} \)
\(D \,-\, {\sj^{\,\pr} \o \sj} \)^{-1}  \, \sj
= D + \( - \sj^{\, 2} + \sj^{\, \pr}\) D^{-1}
\lab{sjlaxred1}
\ee
The last equality was obtained by the trick of multiplying left hand side
by $1= \sj^{\, -1} \sj\,$ from both sides.
Taking however the equivalent Lax operator as given in \rf{sjlax1}
we get automatically again
\be
l_{\sj} = L_{\sj} \bv_{\theta_1 =0} = D - \sj \, D^{-1} \sj
\lab{sjlaxred2}
\ee
Hence the mKdV hierarchy is given in terms of three alternative and equivalent
Lax operators given in \rf{mkdvlax}, \rf{sjlaxreda} and \rf{sjlaxred1}.
Especially the mKdV Hamiltonians (including those in \rf{mkdvham}) are
invariant under transformation $ \sj \, \to - \sj \,$.
\lskip
{\sl Miura Map.}
Take now the generalized Miura transformation \rf{gmiura} and impose the Dirac
constraint $J = - \sj \,- \bsj \, + \sj^{\,\pr} /\sj \, =0 $.
As a result we get the conventional Miura map:
\be
\bj\bv_{J=0} = \sj \, \( - \sj \, + {\sj^{\,\pr}  \o \sj} \) =
-\sj\,^2 + \sj^{\,\pr}
\lab{miura}
\ee
It is easy to find via Dirac procedure that $\sj\,$ satisfies the bracket
\be
\{ \sj \, (x) \, , \, \sj \, (y) \}_2^{D} =
- \int dz dz^{\pr} \{ \sj \, (x) \, , \, J (z) \}_2
\{ J (z) , J (z^{\pr} ) \}_2^{-1}
\{ J (z^{\pr}) \, , \, \sj \, (y) \}_2 = - \h \, \d^{\pr} (x-y)
\lab{mdira3}
\ee
which is perfectly consistent with $ \bj =-\sj\,^2 + \sj^{\,\pr}$ satisfying
the bracket \rf{dira2}.

Especially we see that all Hamiltonians from \rf{kdvham} go to Hamiltonians
in \rf{mkdvham} under $\bj \to - \sj\,^2 \pm \sj^{\,\pr}$.
\lskip
{\sl Bi-Hamiltonian Structure of KdV Hierarchy}
The evolution equation \rf{diraham} specified to the constraint manifold
$ J=0$ results in
\be
\partder{\bj}{t_r} \bv_{J=0} = {\pbr{\bj}{H^{KdV}_r}}^D_2 =
\( D \bj + \bj D + \h D^3 \) \funcder {H_r}{\bj}\bv_{J=0}
\lab{kdvbra2}
\ee
in which one recognizes the second Hamiltonian structure of KdV hierarchy. To
recover the first Hamiltonian structure of KdV hierarchy we recall that
from \rf{iflow} we have for the \faa hierarchy:
\be
\partder{}{t_r} { J \choose \bj} = P_{J\,1} {\d H_{r+1}/\d J\choose
\d H_{r+1}/ \d \bj } = P_{J\,2} { \d H_{r}/ \d J \choose
\d H_{r}/ \d \bj}
\lab{jbjtr}
\ee
where in the last identity we used $P_{J\,1}$ and $P_{J\,2}$ from
\rf{Jp1Jp2}.
Let us now take $r$ odd so $H_{r \pm 1} \to 0$ for $J \to 0$.
We find from \rf{jbjtr} using $P_{J\,1}$ that
\be
\partder{\bj}{t_r} \bv_{J=0} = - D \funcder {H_{r+1}}{J} \bv_{J=0}
\lab{bjtr}
\ee
On the other hand using both $P_{J\,1}$ and $P_{J\,2}$
to calculate $\pa{J}/ \pa {t_{r+1}} $ we find in the general
case
\be
2 D \funcder {H_{r+1}}{J} + \( D^2 + D J \)
\funcder {H_{r+1}}{\bj} = - D \funcder{H_{r+2}}{\bj}
\lab{jtr1}
\ee
However in the limit $J\to 0$ since $H_{r+1} \to 0$ we have also
$\d H_{r+1}/ \d \bj \to 0$.
Note however that we can not claim that also $\d H_{r+1}/ \d J \to 0$
follows in this case.
Therefore summarizing we find
\br
\partder{\bj}{t_r} \bv_{J=0} \eq - D \funcder {H_{r+1}}{J} \bv_{J=0}
= \h D \funcder {H_{r+2}}{\bj} \bv_{J=0}
\nonu \\
\eq \h D \funcder {H^{KdV}_{r+2}}{\bj}
= \(\, \h D^3 + \bj D + D \bj\, \) \funcder {H^{KdV}_r}{\bj} \lab{bihamkdv}
\er
which reproduces well-known result about bi-Hamiltonian structure of KdV
(see also \ct{BAK85}).
Equation \rf{bihamkdv} can be also treated as a recurrence
relation which proves that the system defined by Lax given in \rf{faa3}
is indeed KdV system to all orders of Hamiltonian function.

One can now find the bi-Hamiltonian structure for the case of mKdV.
First we recall a formula \ct{wilson}:
\be
\( D \mp 2 \sj \, \) D \( D \pm 2 \sj \, \)= 2 \( \h D^3 + ( -\sj^{\,2} \pm
\sj^{\,\pr}) D + D ( -\sj^{\,2} \pm \sj^{\,\pr}) \)
\lab{wilson}
\ee
{}From Miura transformation we find \ct{wilson}
\be
\funcder {H^{mKdV}_{r}}{\sj}\, = { D \bj \o D \sj}
\funcder {H^{KdV}_{r}}{\bj} = \( -D - 2 \sj \, \) \funcder {H^{KdV}_{r}}{\bj}
\lab{frechet}
\ee
We therefore have:
\br \lefteqn{
\funcder {H^{mKdV}_{r+2}}{\sj}\, = \( -D - 2 \sj \, \)
\funcder {H^{KdV}_{r+2}}{\bj} } \nonu \\
\eq \( -D - 2 \sj \, \) D^{-1} 2 \( \h D^3 + ( -\sj^{\,2} +
\sj^{\,\pr}) D + D ( -\sj^{\,2} +\sj^{\,\pr}) \)
\funcder {H^{KdV}_{r}}{\bj}\nonu \\
\eq \( -D - 2 \sj \, \) D^{-1} \( D - 2 \sj \, \) D \( D + 2 \sj \, \)
\( -D - 2 \sj \, \)^{-1} \funcder {H^{mKdV}_{r}}{\sj} \nonu \\
\eq \( D +2 \sj \, \) D^{-1} \( D -2 \sj \, \) D\funcder {H^{mKdV}_{r}}{\sj}
= \( D - 4 \sj \,D^{-1} \sj\, \) D \funcder {H^{mKdV}_{r}}{\sj}
\lab{frecheta}
\er
where we used both \rf{bihamkdv} and \rf{wilson}.
Relation \rf{frecheta} reveals a bi-Hamiltonian (but non-local) structure
of mKdV hierarchy and can be rewritten in a more simple way as Lenard's
recursion relation:
\be
D \, \funcder {H^{mKdV}_{r+2}}{\sj}\, =
\( D^3 - 4 D \, \sj \,D^{-1} \sj\,D\, \) \, \funcder {H^{mKdV}_{r}}{\sj}
\lab{mkdvbiham}
\ee
\lskip
{\sl Schwarzian-KdV Hierarchy.}
Here few remarks are given about Schwarzian-KdV (S-KdV) hierarchy.
We start be discussion of invariance of the Miura map
$ \bj = -\sj^{\, 2} + \sj^{\, \pr} =  -(\p^{\pr})^{\, 2} + \p^{\pr \pr}$
where as before $ \p^{\pr} = \sj\,$.
Invariance of $\bj$ under some transformation $\d$ results in
\be
\d \(  -(\p^{\pr})^{\, 2} + \p^{\pr \pr} \) = 0 \quad \to \quad
\d \p^{\pr \pr} = 2 \p^{\pr}\d \p^{\pr}
\lab{invariance}
\ee
Solution of \rf{invariance} takes therefore a simple form
\be
\d \p^{\pr} = \d \sj \, = \eps^{-1} \exp ( 2 \p )
\lab{dppr}
\ee
or
\be
\d \p = {\eps^{0}\o 2} + \int \eps^{-1} \exp ( 2 \p )
\lab{dp}
\ee
where $\eps^{0}$ and $ \eps^{-1} $ are some arbitrary constants.
Introduce now the function $f$ such that $ f^{\pr} = \exp (2 \p)$
and which is connected to $\sj$ through the Cole-Hopf type of transformation
\be
\sj \, = \p^{\pr} = \h { f^{\pr \pr} \o f^{\pr}}
\lab{cole}
\ee
We find that \rf{dp} corresponds to $sl_2$ transformation
\be
\d f = \eps^{1}+ \eps^{0} f + \eps^{-1} f^2
\lab{dpf}
\ee
and leaves $ \bj = S (f)/2$ invariant, where $S(f)$ is a Schwarzian.

It is known that \rf{cole}
relates the mKdV hierarchy to the S-KdV hierarchy with equation
$f_t /f^{\pr} = S(f)$. We are using Weiss
nomenclature \ct{weiss}, \ct{fuchs} is using the name of KdV-singularity
hierarchy.
Hence according to \rf{cole} we will be interested in one-boson Lax operator
of the form
\be
L= D - \h { f^{\pr \pr} \o f^{\pr}} D^{-1} \h { f^{\pr \pr} \o f^{\pr}}
\lab{schwarzlax}
\ee
There are many ways of promoting this operator to two-boson system.
If we consider a very simple choice
\be
L= D + \h { f^{\pr \pr} \o f^{\pr}} + \sj \,+ { f^{\pr \pr} \o f^{\pr}}
D^{-1} \sj
\lab{schwarzlax1}
\ee
the second bracket structure is
\be
\pbr{\sj\,(x)}{f(y)} = -2 f^{\pr} (x) D^{-1}_x \d (x-y)
\lab{sjfbra}
\ee
Another choice could be
\be
L= D + { f^{\pr \pr} \o f^{\pr}} + 2 \rho + \({ f^{\pr \pr} \o f^{\pr}}+
\rho\) D^{-1} \rho
\lab{schwarzlax2}
\ee
leads to \rf{schwarzlax} under constraint ${ f^{\pr \pr} \o f^{\pr}} +
2 \rho =0$.
Defining $ \rho = v^{\pr}$ we can now make contact with quadratic KP hierarchy
by defining a map:
\be
\bsj \, = v^{\pr} + { f^{\pr \pr} \o f^{\pr}} \quad ; \quad
\sj \, =  v^{\pr}
\lab{schwarzmiura}
\ee
Of course the ambiguity of \rf{sjtransf} allows equally well
the map:
\be
\bsj \, = v^{\pr} + { f^{\pr \pr} \o f^{\pr}}-
{ v^{\pr \pr} \o v^{\pr}} \quad ; \quad \sj \, =  v^{\pr}
\lab{schwarzmiura1}
\ee
The structure in \rf{schwarzmiura} has a non-local bracket structure
equivalent to structure in \rf{p2sj}. We find easily e.g.
\be
\pbr{v(x)}{f(y)} = D^{-1}_x f^{\pr} (x) D^{-1}_x \d (x-y)
\lab{vfbra}
\ee
\sect{Two-boson KP Hierarchies in Terms of a SL(2) Gauge Theory}
\subsection{Zero Curvature Condition and Soliton Equations.}
We first establish a connection between a typical two-boson Lax operator $L$
of \mKPa hierarchy characterized by three functions $A,B,C$
and a component $\ca$ of $SL (2, \IR)$ Lie algebra valued gauge field.
We express this connection in terms of the following equivalence relation:
\be
L = D + A +B \, D^{-1} \, C \; \; \sim \; \;
{\cal A} = \fourmat{-\h A}{-C}{B}{\h A}
\lab{laxabc}
\ee
Under the gauge transformation applied on $L$ the above equivalence takes the
following form
\be
L^{\pr}  = e^{-\chi}L e^{\chi}= D + (A+ \pa \chi) +
(e^{-\chi} B) \, D^{-1} \, (C e^{\chi}) \; \; \sim \; \;
{\cal A}^{\pr} = \fourmat{-\h (A+\pa \chi)}{(C e^{\chi})}{(e^{-\chi} B)}
{\h (A+\pa \chi)}
\lab{laxprabc}
\ee
We note that by the above equivalence principle a gauge transformation
among Lax operators of \mKPa corresponds to the $SL (2, \IR)$
gauge transformation:
\be
{\cal A}^{\pr} = g {\cal A} g^{-1} + g \pa g^{-1}
\lab{sl2transf}
\ee
induced by the following diagonal $2\times2$-real
unimodular matrix:
\be
g \equiv \fourmat{e^{\chi/2}}{0}{0}{e^{-\chi/2}}
\lab{gchi}
\ee
One easily verifies that the following three gauge configurations
\be
\fourmat{-\h A}{-C}{B}{\h A} \sim \fourmat{-\h(A+B^{\pr}/B)}{-BC}{1}
{\h(A+B^{\pr}/B)} \sim
\fourmat{-\h(A- C^{\pr}/C)}{-1}{BC}{\h(A-C^{\pr}/C)}
\lab{configs}
\ee
are gauge equivalent with gauge functions
\be
g_{B} = \fourmat{B^{\h}}{0}{0}{B^{-\h}} \qquad ; \qquad
g_{C} = \fourmat{C^{-\h}}{0}{0}{C^{\h}}
\lab{gbc}
\ee
generating connections between the first and the second and the first and the
third gauge connection of equation \rf{configs}.

Let us recall three main examples of the two-boson \mKPa hierarchies
with their corresponding components of the $sl (2, \IR)$ connection:
\br
L_{NLS} \eq D + {\bar \psi} D^{-1} \psi \, \sim \,
\ca_{NLS} = \fourmat{0}{-\psi}{{\bar \psi}}{0}  \lab{nls}\\
L_{\sj\,} \eq D + \sj \,+ \bsj\, + \bsj \, D^{-1} \sj \, \sim \,
\ca_{\sj\,} = \fourmat{- \h (\sj\,+\bsj\,)}{-\sj\,}{\bsj\,}{\h (\sj\,+\bsj\,)}
\lab{mKP2b}\\
L_{J} \eq D - J + \bj D^{-1} \, \sim \,
\ca_{J} = \fourmat{\h J}{-1}{\bj}{-\h J}   \lab{KP2b}
\er
Defining the element of $SL (2, \IR)$:
\be
g_{\psi} = \fourmat{\psi^{-\h}}{0}{0}{\psi^{\h}}
\lab{gpsi}
\ee
we are able to transform $\ca_{NLS}$ to the form of $\ca_{J}$:
\be
\ca_{NLS}^{\pr} = g_{\psi} \ca_{NLS} g^{-1}_{\psi} + g_{\psi}
\pa g^{-1}_{\psi} = \fourmat{\h \pa_x \psi/\psi}{-1}{\psi {\bar \psi}}
{-\h \pa_x \psi/\psi}
\lab{nlspr}
\ee
Similarly applying gauge transformation generated by
\be
g_{\sj\,} = \fourmat{e^{-\h (\p +\bp)}}{0}{0}{e^{\h (\p +\bp)}}
\lab{gsj}
\ee
to $\ca_{\sj\,}$ we get
\be
\ca_{\sj\,}^{\pr} = g_{\sj\,} \ca_{\sj\,} g^{-1}_{\sj\,} + g_{\sj\,}
\pa g^{-1}_{\sj\,} = \fourmat{0}{-e^{-\p -\bp}\sj\,}{\bsj\,e^{\p +\bp}}
{0}
\lab{mkppr}
\ee
the gauge field component belonging to NLS hierarchy.

We now put the $\ca$ component in the complete the $sl(2,\IR)$ connection
$\( \ca , \cb \)$ satisfying the zero curvature condition \ct{chern}:
\be
\pa_x \cb - \pa_t \ca + \sbr{\cb}{\ca} = 0   \lab{curv0}
\ee
where we have introduced the other component of the $sl (2, \IR)$
gauge field:
\be
\cb = \fourmat{B^0}{B^+}{B^-}{-B^0}
\lab{bmatrix}
\ee
In components \rf{curv0} reads:
\br
\pa_x B^0 -\psi \, B^-  - {\bar \psi} B^+ \eq 0 \lab{curva}\\
\pa_x B^+ + \pa_t \psi + 2 \psi \, B^0 \eq 0 \lab{curvb}\\
\pa_x B^- - \pa_t {\bar \psi} + 2 {\bar\psi} \, B^0 \eq 0 \lab{curvc}
\er
Taking
\be
\cb_{NLS} = \fourmat{\psi {\bar \psi}}{\pa_x \psi}{\pa_x {\bar \psi}}
{- \psi {\bar \psi}}
\lab{bnls}
\ee
we see that \rf{curva} is satisfied automatically while \rf{curvb}
and \rf{curvc} yield equations of NLS hierarchy \ct{mason}:
\be
\pa_t {\psi} = - \pa_x^2 \psi - 2 \psi^2 {\bar \psi} \quad;\quad
\pa_t {\bar {\psi}} = \pa_x^2 {\bar \psi} +2 \psi {\bar \psi}^2
\lab{nlseqs}
\ee

It is interesting at this point to comment on
connection between NLS system and the Heisenberg Model.

Since $ \ca_{NLS}$ and $\cb_{NLS}$ satisfy the zero curvature equation
\rf{curv0} it is natural to represent them (locally)
as pure gauge configurations \ct{ZakTak}:
\br
\ca_{NLS} &=& g^{-1} \, g_{x} \nonu\\
\cb _{NLS} &=& g^{-1} \, g_{t}
\lab{g}
\er
Introduce now the traceless matrix:
\be
S= g\sigma_3g^{-1}
\lab{S}
\ee
which has the property $S^2 = 1$.
It can easily be shown that:
\be S \, S_x = -S_x \, S = -2g_x \, g^{-1}
\lab{sx}
\ee
and consequently
\be
g^{-1}[S,S_{xx}]g = -4 {d \o {dx}}\ca _{NLS}
\ee
On the other hand, taking time derivative of $S$ we find
\be
g^{-1}S_tg = \sbr{\cb_{NLS}}{\sigma_3}
 \ee
Using explicit form of $\ca _{NLS} $ and $ \cb _{NLS}$ given above
in \rf{nls} and \rf{bnls}, we arrive at
\be
S_t = \h \, \sbr{S}{S_{xx}}
\ee
which describes the isotropic Heisenberg ferromagnet model

Applying the gauge transformation \rf{gpsi} to $\cb_{NLS}$ we obtain
\be
\cb_{J} = \fourmat{\bj + \h \pa_t \Phi}{J}{\pa_x \bj -\bj J}
{- \bj - \h \pa_t \Phi}
\lab{bkp}
\ee
which when inserted into \rf{curv0} (together with $\ca_{J}$)
yields the Bussinesq equations.

\subsection{Reduction to the KdV Systems.}
It is also easy in this framework to discuss the reduction of KP systems
to KdV systems.
The link is obtained by putting $\sj\,+\bsj\,=0$ and $J=0$ in
\rf{mKP2b} and \rf{KP2b} getting
\be
\ca_{mKdV} = \fourmat{0}{-\sj\,}{-\sj\,}{0}
\quad;\quad
\ca_{KdV} = \fourmat{0}{-1}{\bj}{0}   \lab{kdvmat}
\ee
Solutions to zero curvature equation involving gauge fields components
of type given in \rf{kdvmat} (with spectral parameter $\l$ instead of zeros)
have been discussed in \ct{chern}.
We recall the main points of this discussion.
Let us first take $\ca_{mKdV}$ modified by adding to it the spectral parameter
$\l\sigma_3$. After inserting it and matrix \rf{bmatrix}
into the zero-curvature equation \rf{curv0} we obtain
\br
\pa_x B^0 - \sj\, b_- \eq 0 \lab{mkdvcurva}\\
\pa_x b_- - 4 \sj\, B^0 - 2 \l b_+ \eq 0 \lab{mkdvcurvb}\\
\pa_x b_+ + 2 \pa_t \sj \, - 2 \l b_- \eq 0 \lab{mkdvcurvc}
\er
where for convenience we have introduced $b_{\pm} \equiv B^+ \pm B^-$.
We now eliminate $b_{\pm} $ in terms of $B^0$ using \rf{mkdvcurva}
and \rf{mkdvcurvb}.
First we introduce the function $C ( \sj\, , \l)$ such that
\be
b_- = \pa_x B^0 / \sj \, = \l  C^{\pr}        \lab{cdef}
\ee
Hence we find that $B^0 = \l D^{-1} \sj \, C^{\pr}$ and from \rf{mkdvcurvb}
we get $ b_+ =- 2 \sj \,D^{-1} \sj \, C^{\pr} + C^{\pr \pr}/2$.
Inserting these quantities into \rf{mkdvcurvc} we arrive at
\be
\pa_t \sj \, = \( D \sj \,D^{-1} \sj \, C^{\pr} - {1 \o 4} C^{\pr \pr\pr} \)
+ \l^2 C^{\pr}
\lab{mkdvjt}
\ee
Expanding $C$ in $\l$ as in $ C = \sum_{k=0}^n (\l^2)^{n-k} C_k$ we obtain
from \rf{mkdvjt}:
\br
D \, C_k \eq \( {1 \o 4} D^3 - D \sj \,D^{-1} \sj \, D \) C_{k+1} \qquad
k = 0 , 1 ,\ldots, n-1  \nonu \\
\pa_t \sj \, \eq \( D \sj \,D^{-1} \sj \, D  - {1 \o 4} D^3 \) C_{n}
\lab{mkdvjta}
\er
We clearly recognize in \rf{mkdvjta} the bi-hamiltonian structure of mKdV
equation. Moreover putting $C_n = \sj \, $ we recover
the mKdV equation
$\pa_t \sj \, = D \sj \,D^{-1} \sj \, D \sj\,  - D^3 \sj\, /4=
3 \sj^{\,2} \sj^{\, \pr} /2 - \sj^{\, \pr \pr \pr} /4$.

Similar results hold in case of KdV represented by
$\ca_{KdV}$ modified by addition of $\l \sigma_3$.
After inserting it and matrix \rf{bmatrix} into the zero-curvature
equation \rf{curv0} and eliminating $B^0$ and $B^-$ we arrive at equation
\be
\pa_t \bj = \( - \h D^3 + D \bj + \bj D \) B^+ \, + \, 2 \l^2 B^+
\lab{kdvjt}
\ee
Also here expanding $B^+$ in powers of $\l$ reveals both bi-Hamiltonian
structure and KdV equation behind the equation \rf{kdvjt}.

\subsection{Drinfeld-Sokolov reduction of
Two-boson Hierarchies to KdV}

As we have seen above we can associate to each two-boson Lax operator
a $sl_2$ matrix according to \rf{laxabc}, in such a way that the gauge
transformation of the Lax operator $L^{\pr}  = e^{-\chi}L e^{\chi}$
corresponds to the $sl_2$ gauge transformation of $sl_2$ connection
${\cal A}^{\pr} = g {\cal A} g^{-1} + g \pa g^{-1}$ with diagonal
$2\times2$-real unimodular matrix \rf{gchi}.

Let us now take the special example of \faa hierarchy represented
by $\ca_{J}$ from \rf{KP2b}.
Important point is that there is a residual gauge transformation generated
by
\be
g_0 \equiv \fourmat{1}{0}{\g}{1}
\lab{resi}
\ee
which preserves the form of $\ca_{J}$ under
\be
{\cal A}^{\pr} = g_0^{-1} {\cal A} g_0 + g_0^{-1} \pa g_0 =
\fourmat{\h J-\g}{-1}{\bj- \g J + \g^2 + \g^{\pr} }{-\h J+\g}
\lab{KPsl2pr}
\ee
It is useful at this point to explain what is happening using
the Drinfeld-Sokolov formalism \ct{DrinSok}.
Consider space of first order differential operators
with coefficients being $2\times 2$ matrices:
\be
M_{\ce} = \lcurl L^{(1)} = D - \ce + \om \bv \ce = \fourmat{0}{1}{0}{0}
\; , \; \om =  \fourmat{\om_{11}}{0}{\om_{21}}{\om_{22}}
\rcurl
\lab{dsm}
\ee
and the group
\be
\G \equiv \lcurl \G \bv \G \equiv \fourmat{1}{0}{\g}{1} \rcurl
\lab{resid}
\ee
acting on $M_{\ce}$ according to
\be
\G^{-1} \( D - \ce + \om \) \G = D - \ce + \om^{\pr}
\lab{dceom}
\ee
with
\be
\om^{\pr} =  \fourmat{\om_{11}-\g}{0}{\om_{21} - \g (\om_{22} - \om_{11})
+ \g^2 + \g^{\pr}}{\om_{22}+\g}
\lab{ompr}
\ee
In the spirit of Hamiltonian Drinfeld-Sokolov reduction consider the
quotient space $M_{\rm red} = M_{\ce} / \G$.
There exist a convenient realization of $M_{\rm red}$ in terms of second order
differential operators with scalar coefficients.
The procedure to obtain it is as follows. Consider the relation
\be
L^{(1)} { \psi_1 \choose \psi_2} =  0
\lab{dspsi}
\ee
Eliminating $\psi_2$ from this equation we arrive at $L^{(2)} \psi_1=0$ with
\be
L^{(2)} = \a \( L^{(1)}\) = D^2 + ( \om_{11} + \om_{22} ) D
+ \om_{21} + \om_{11} \om_{22} + \om_{11}^{\pr}
\lab{l2}
\ee
Because $\a \( \G^{-1} L^{(1)} \G\) = \a \( L^{(1)}\)$ the space of second
order differential operators from \rf{l2} parameterizes the quotient space
$M_{\rm red} $.

Consider now the special case of two-boson KP hierarchy:
\be
\om = \fourmat{\om_{11}}{0}{\om_{21}}{\om_{22}} =
\fourmat{\h J}{0}{\bj}{-\h J}
\lab{omj}
\ee
Take $\G$ with $\g= \h J$ so the transformed $\om$ matrix
\be
\om^{\pr} = \fourmat{0}{0}{u }{0}
= \fourmat{0}{0}{\bj- {1\o 4} J^2 + \h J^{\pr} }{0}
\lab{diagom}
\ee
has diagonal elements equal to zero. It means that the corresponding
Lax operator is:
\be
L^{(1)} = D + u D^{-1} \qquad;\qquad
u = \bj- {1\o 4} J^2 + \h J^{\pr}
\lab{udef}
\ee
One can check that with $(\bj , J)$ satisfying the second Poisson bracket
\rf{33} with $c=2, h=1$, $u$ satisfies the Virasoro algebra:
\be
\pbr{u(x)}{u(y)} = 2 u (x)\, \d^{\pr} (x-y) + u^{\pr} (x) \,\d (x-y) +
\h \d^{\pr \pr \pr} (x-y)
\lab{uvira}
\ee
We also note that with $\om$ like in \rf{omj} the second-order
differential operator \rf{l2} becomes a typical KdV operator $L^{(2)} =
D^2 + u$.
Hence $\om^{\pr}$ from \rf{diagom} or first order Lax $L^{(1)}$
from \rf{udef} represent just the special gauge choice on $M_{\ce}$
equivalent to the KdV Lax operator.
This shows Drinfeld-Sokolov reduction as an alternative to the Dirac
reduction of two-boson hierarchy to KdV hierarchy.

\end{document}